\documentclass[twocolumn,aps]{revtex4-2}
\usepackage[latin9]{inputenc}
\usepackage{textcomp}
\usepackage{amsmath}
\usepackage{amssymb}
\usepackage{graphicx}
\usepackage{makecell}
\usepackage{float}
\usepackage{hyperref}
\usepackage{caption}
\usepackage{subcaption}

\makeatletter

\DeclareFontEncoding{LGR}{}{}

\ProvideTextCommand{\~}{LGR}[1]{\char126#1}

\@ifundefined{textcolor}{}{%
\definecolor{BLACK}{gray}{0}
\definecolor{WHITE}{gray}{1}
\definecolor{RED}{rgb}{1,0,0}
\definecolor{GREEN}{rgb}{0,1,0}
\definecolor{BLUE}{rgb}{0,0,1}
\definecolor{CYAN}{cmyk}{1,0,0,0}
\definecolor{MAGENTA}{cmyk}{0,1,0,0}
\definecolor{YELLOW}{cmyk}{0,0,1,0}
}

\usepackage{upgreek}
\usepackage{braket}

\usepackage{hyperref} 
\usepackage{xr-hyper}

\providecommand{\tabularnewline}{\\}

\@ifundefined{textcolor}{}{%
\definecolor{BLACK}{gray}{0}
\definecolor{WHITE}{gray}{1}
\definecolor{RED}{rgb}{1,0,0}
\definecolor{GREEN}{rgb}{0,1,0}
\definecolor{BLUE}{rgb}{0,0,1}
\definecolor{CYAN}{cmyk}{1,0,0,0}
\definecolor{MAGENTA}{cmyk}{0,1,0,0}
\definecolor{YELLOW}{cmyk}{0,0,1,0}
}

\usepackage{color}%\bibliographystyle{plain}
\usepackage{xcolor}
\def\NOT(#1,#2){\OneQubitGate(#1,#2){$X$}}

\@ifundefined{showcaptionsetup}{}{%
\PassOptionsToPackage{caption=false}{subfig}}

\makeatother

\begin{document}

\title{Towards a spectrally multiplexed quantum repeater
}

% Frequency multiplexed photon pairs and spectrally resolved detection assisted by a thulium-doped spectral filter for quantum repeaters

%% Original title:
%\title{Frequency multiplexed photon pairs and detection for quantum repeaters 
%\WT{?Frequency multiplexed photon pairs, quantum memory-like spectral filtering, and spectrally resolved detection for quantum repeaters?}}

%% 

%\footnote[$^{\dagger}$]{Current affiliation: Netherlands Organisation for Applied Scientific Research (TNO), P.O. Box 155, 2600 AD Delft, Netherlands}
 
\author{Tanmoy Chakraborty$^{1}$}

\author{Antariksha Das$^{1}$}

\author{Hedser van Brug$^{2}$}

\author{Oriol Pietx-Casas$^{1}$}

\author{Peng-Cheng Wang$^{1}$}

\author{Gustavo Castro do Amaral$^{1*}$}

\author{Anna L. Tchebotareva$^{1,2}$}

\author{Wolfgang Tittel$^{1,3,4}$}

\affiliation{$^{1}$QuTech, and Kavli Institute of Nanoscience, Delft University of Technology, 2628 CJ Delft, The Netherlands}
\affiliation{$^{2}$Netherlands Organisation for Applied Scientific Research (TNO), P.O. Box 155, 2600 AD Delft, The Netherlands}
\altaffiliation{Current address: Netherlands Organisation for Applied Scientific Research (TNO), P.O. Box 155, 2600 AD Delft, The Netherlands.}

\affiliation{$^{3}$Department of Applied Physics, University of Geneva, 1211 Geneva 4, Switzerland}
\affiliation{$^{4}$Constructor University, 28759 Bremen, Germany}

\begin{abstract}
Extended quantum networks are based on quantum repeaters that often rely on the distribution of entanglement in an efficient and heralded fashion over multiple network nodes. Many repeater architectures require multiplexed sources of entangled photon pairs, multiplexed quantum memories, and photon detection that distinguishes between the multiplexed modes. Here we demonstrate the concurrent employment of (1) spectrally multiplexed cavity-enhanced spontaneous parametric down-conversion in a nonlinear crystal; (2) a virtually-imaged phased array that enables mapping of spectral modes onto distinct spatial modes for frequency-selective detection; and (3) a cryogenically-cooled Tm$^{3+}$:LiNbO$_3$ crystal that allows spectral filtering in an approach that anticipates its use as a spectrally-multiplexed quantum memory. Through coincidence measurements, we demonstrate quantum correlations between energy-correlated photon pairs and a strong reduction of the correlation strength between all other photons. This constitutes an important step towards a frequency-multiplexed quantum repeater.

\end{abstract}

\maketitle

%\textcolor{red}{check references}

\emph{Introduction.} Triggered by many proof-of-principle experiments over the past decades \cite{gisin2002quantum,lo2014secure,xu2020secure, clivati2022coherent,wang2022twin}, quantum key distribution (QKD) has reached a level of maturity that allows building large-scale networks over standard telecommunication fiber \cite{chen2021implementation,dynes2019cambridge}. However, due to absorption of photons, long-distance transmission still requires the use of trusted nodes \cite{peev2009secoqc}, even though QKD has recently been reported over 1000 km using an idealized laboratory setting with spooled fiber \cite{Liu2023}.  
To enable information-theoretic secure QKD and entanglement-based applications, these trusted nodes have to be replaced by (untrusted) quantum repeaters, which promise the creation of distant entanglement with improved scaling compared to direct transmission \cite{takeoka2014fundamental,pirandola2017fundamental}. 

In many quantum repeater schemes, long-lived entanglement is established between multi-mode quantum memories positioned at the opposite ends of ``elementary links" of limited length, e.g. 10-100 km \cite{sangouard2011quantum}. Each attempt is probabilistic, but the use of a sufficiently large number of modes ensures success with large probability in at least one of them \cite{sinclair2014spectral}-- which one being indicated by a ``heralding signal". In turn, this allows extending entanglement by means of entanglement swapping across neighboring elementary links after photons have been retrieved from their memories and their modes shifted to make them indistinguishable. Such feed-forward control has been implemented using various degrees-of-freedom (DOF) of the electromagnetic field%, including temporal, spectral, spatial and orbital angular momentum modes 
\cite{Ortu2021,sinclair2014spectral,Lan2009,VernazGris2018,yang2018multiplexed,wang2021long,pu2017experimental}.

\begin{figure*}
\includegraphics[width=1.6\columnwidth]{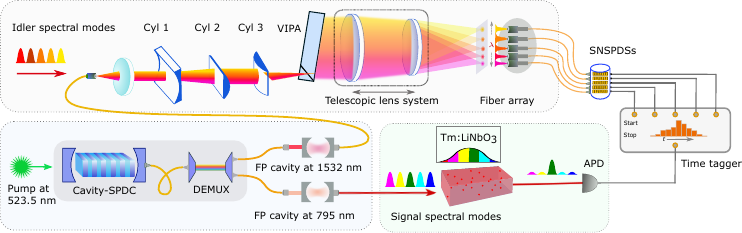}
\caption{Schematic of the experimental setup including a spectrally-multiplexed photon pair source, a VIPA-based demultiplexer for mapping spectral modes to distinct spatial channels, and a Tm$^{3+}$:LiNbO$_3$-based spectral filter: APD - avalanche photodiode; DDG - digital delay generator; SNSPD - superconducting nanowire single photon detector; AD - achromatic Doublet (focusing lens coupling the output beam into collection fiber). CyL1, CyL2 and CyL3 are cylindrical lenses with $f$=-50 mm, 200 mm and 150 mm, respectively.
\label{fig:SPDC_VIPA}}
\end{figure*}

To enable the required multi-mode operation of an elementary link, all its elements---sources of entangled photon pairs, optical quantum memory, and single-photon detectors---must allow, respectively, emitting, storing, and detecting photons in a multi-mode or mode-selective manner. However, while the creation, storage and detection of temporally multiplexed photon pairs is common, the other DOFs have received much less attention, and their practical value---including the joint operation of all components---remains to be assessed. This is particularly important for spectral modes since their use simplifies the quantum memory---no readout-on-demand is required \cite{sinclair2014spectral}---but, at the the same time, imposes new constraints on sources and detectors.

In 2014, Sinclair \textit{et al.} proposed a repeater scheme that exploits frequency multiplexing, and demonstrated spectrally multimode photon storage in a cryogenically-cooled Tm$^{3+}$:LiNbO$_3$ crystal together with mode shifting after feed-forward control \cite{sinclair2014spectral}. To complement this work, here we focus on the demonstration of the multiplexed photon pair source, demultiplexed detection of photons in different spectral modes, and, as a precursor to frequency-multiplexed quantum memory, a programmable spectral filter based on the same crystal. More precisely, we implement a novel alignment-free, frequency multiplexed photon pair source that is based on fiber-pigtailed, cavity-enhanced spontaneous parametric down-conversion (SPDC) in a nonlinear crystal and is easy to integrate within a practical quantum repeater. Our work builds on previous demonstrations of cavity-enhanced SPDC \cite{Ou1999,Scholz2007,pomarico2009waveguide,fekete2013ultranarrow,fortsch2013versatile,Ikuta2019,Seri2019}, but extends them from the characterization of either a single pair of spectrally-resolved modes or a large number of unresolved modes, to many pairs of spectrally-resolved modes, as required for a quantum repeater. 

Furthermore, as a simple spectral filter does not allow one to distinguish between different spectral modes (only one mode can be selected at a time), we also study a novel approach to frequency-demultiplexed photon detection based on a Virtually Imaged Phased Array (VIPA)  \cite{shirasaki1996large,shirasaki1999virtually,xiao2005eight, pietx2023spectrally} connected to an array of 8 fibers out of which we use 5. The demultiplexer provides a system efficiency up to 17\%, superior spectral resolution compared to standard diffraction gratings  \cite{puigibert2017heralded}, and cross-talk below $-25\,$dB. See the Supplemental Materials (SM) for details.
 
Finally, to filter the $795\,$nm photons, we create a spectral trench through persistent spectral hole burning \cite{macfarlane2002}. This approach is closely related to atomic-frequency-comb (AFC)-based quantum memory for light \cite{saglamyurek2011broadband}, differing only in the creation of a spectral bin with uniformly low optical depth instead of a periodic modulation (an AFC). %Note that the repeater protocol in \cite{sinclair2014spectral} requires only a single spectral channel at this wavelength.
Coincidence measurements reveal non-classical correlations between spectrally correlated photon pairs, and strongly reduced correlations between photons belonging to non-matched spectral channels.

\emph{Setup.} Fig.\ref{fig:SPDC_VIPA} depicts a schematic of the experimental setup, and its integration into a quantum repeater is described in the SM. %\ref{fig:Suppl_fig1.pdf}}.%, along with its integration into a spectrally-multiplexed repeater. %is depicted in the Supplemental Material (SM). 
To create spectrally multiplexed photon pairs, we exploit spontaneous parametric down-conversion in a $1\,$cm long, type-0, quasi-phase-matched, periodically poled LiNbO$_3$ (PPLN) crystal waveguide. At a temperature of $44.5^{\circ}$C, the poling period of $6.9\,\mu$m enables the interaction between $523.5\,$nm continuous-wave pump light and ``signal" and ``idler" photons with spectra centered at $795\,$nm and $1532\,$nm, respectively. To enable future use of this source in a quantum repeater, the signal wavelength matches the absorption line of Tm-doped crystals \cite{thiel2014tm,sinclair2014spectral,askarani2021long}, which are promising platforms for spectrally multiplexed quantum storage. Furthermore, the idler photon wavelength of $1532\,$nm allows low-loss transmission through fiber networks.

The PPLN is part of a monolithic Fabry-P\'erot cavity, created by reflection coatings at its input and output facets ($R\,\approx\,99\%$ at $1532\,$nm, T$\,>\,$99\% at $523.5\,$nm and $795\,$nm, HC Photonics Corp.).  
The cavity resonances restrict the spectrum of the idler photons to discrete spectral intervals. In turn, energy conservation also modifies the signal spectrum. Indeed, assuming a coherent pump laser with frequency $\omega_p$, we find that $\omega_s \,=\, \omega_p-\omega_i$ (where $s$, $i$, label  signal and idler photons, respectively), and thus perfect energy correlations between pairs of signal and idler spectral modes. Note that the cavity is singly resonant at $1532\,$nm to avoid clustering \cite{luo2015direct,Seri2019}.  

To ease the use of the SPDC cavity in a practical setting, 
its input is pigtailed to a single-mode fiber for $523$\,nm wavelength, and the output is fiber-coupled to a 1:2 demultiplexer that separates signal and idler photons into two separate fiber pigtails. The rate of emitted $1532\,$nm photons, calculated from detection rates, is around 6 times smaller than that of the $795\,$nm photons, suggesting a misaligned pigtail. After filtering out all residual pump light, the width of the signal and idler spectra are reduced using frequency tunable Fabry-P\'erot \'etalons of $6.1\,$GHz and $16\,$GHz bandwidths, respectively.

The $795\,$nm signal photons are directed to a spectral filter based on a trench in the absorption profile of a Tm$^{3+}$:LiNbO$_3$ crystal cooled to $\approx\,$600 mK. This trench is created by means of persistent spectral hole-burning, i.e. the transfer of Tm ions using a $795\,$nm laser over a $100\,$MHz bandwidth from the ground state to an excited state. From the excited state, ions subsequently decay spontaneously to a long-lived state---a second ground-state that arises under the application of a 1.2 Tesla magnetic field---that does not interact with the laser light anymore. A typical trench is depicted in Fig. \ref{fig:g2_map}c. The central frequency of this spectral filter can easily be tuned by changing the wavelengths over which the laser is swept during hole burning. The filtered photons are subsequently detected using a Silicon avalanche photodiode (APD)-based single-photon detector featuring a detection efficiency of 55$\%$, a dark count rate of around 60 Hz, and approx. 600 ps detection time jitter.

To spectrally demultiplex the idler modes at around $1532\,$nm, we use a VIPA (Light Machinery Inc.) with resolving power $\lambda /\Delta \lambda\simeq 2.5 \times 10^5$, free spectral range (FSR) of $\sim 60.8\,$GHz and bandwidth of $\sim 0.76\;$GHz (FWHM). The input light exits a single-mode fiber via a series of beam shaping lenses, and is focused onto the entrance slit of the VIPA.
A telescopic lens system with an effective focal length of 60$\,$mm maps the exiting spectral modes onto distinct spatial modes that are matched to the position and the numerical aperture of five standard single-mode fibers in a fiber array (PHIX Photonics Assembly). For the initial characterisation of the VIPA we use a single collection fiber instead of the fiber array, and a lens with fixed focal length of $45\,$mm instead of the telescopic system. See SM for more information. The idler photons are detected by WSi Superconducting Nanowire Single Photon Detectors (SNSPDs) featuring detection efficiencies around 65$\%$, dark count rates around $70\,$Hz, and approx. $250\,$ps detection time jitter.

Finally, the signals from the single-photon detectors---one for the $795\,$nm photons and five for those at $1532\,$nm---are sent to a time tagger that outputs coincidence detection rates.

\begin{figure}
\includegraphics[width=0.85\columnwidth]{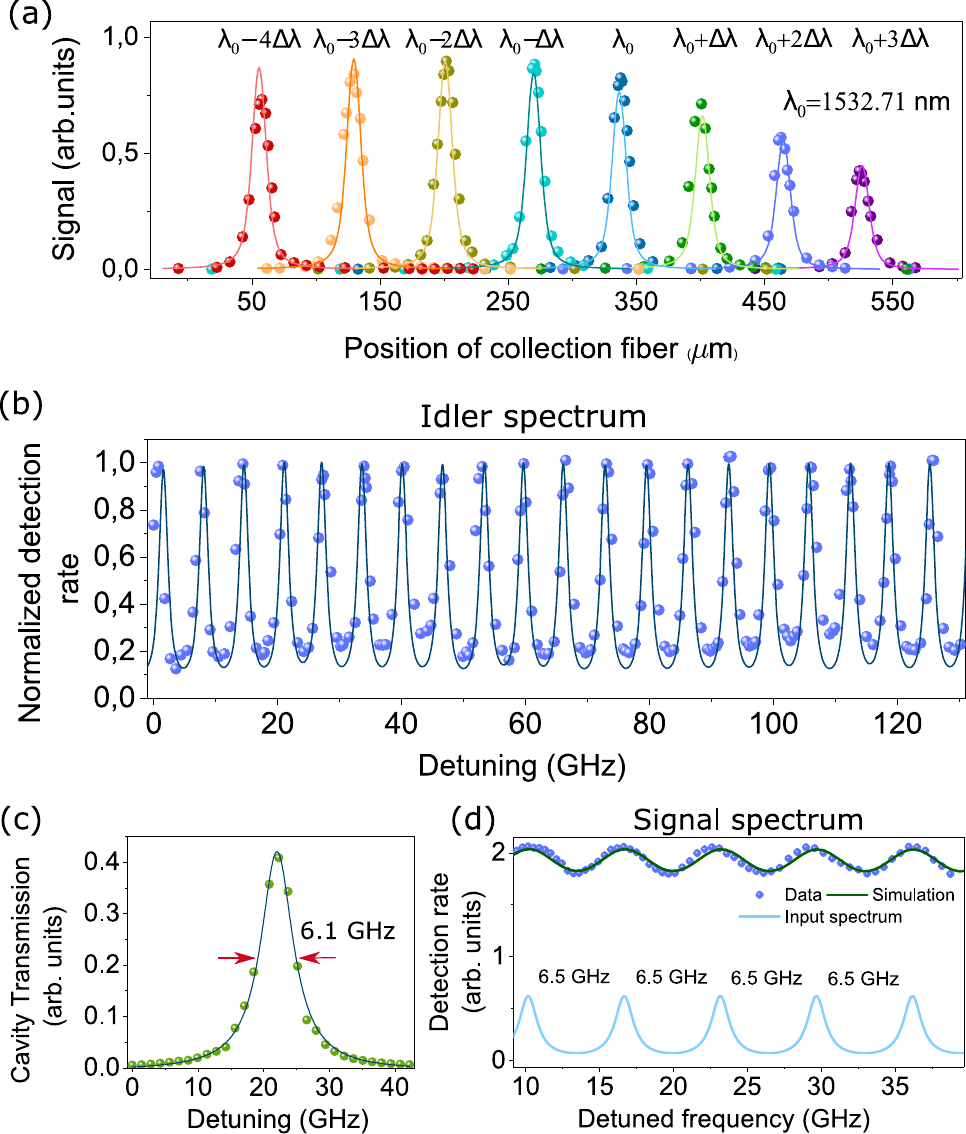}
\caption{Spectroscopy using the VIPA. (a) Measured signal intensity as a function of the position of the VIPA output fiber for eight spectral modes separated by $\Delta\nu=6.5\,$GHz (dots) created by shifting the laser frequency, and  simulation results (solid lines) (b) Spectrum for twenty idler modes created by the SPDC source (dots) as measured using the VIPA setup, and simulation results (solid line). (c) Transmission profile of the Fabry-P\'erot filter cavity operating in the range of 795 nm (dots: measured data; solid line: fit). (d) Spectrum of the 795 nm signal photons measured by detuning the $6.1\,$GHz filter cavity (dots), simulated frequency comb with $6.5\,$GHz mode spacing (light blue line), and its convolution (dark blue line) with the cavity profile shown in (c) (green curve). 
\label{fig:spectra}}
\end{figure}

\emph{Measurements and results.} As a first step, we demonstrate the demultiplexing capability of our VIPA-setup using a tunable continuous-wave laser, a single collection fiber, and a linear photodetector instead of the photon pair source, the fiber array, and the SNSPDs, respectively. For light at $\lambda_0\,=\,1532.71\,$nm, we couple the brightest order of the VIPA into the collection fiber, and measure the transmitted intensity as a function of laser detuning. We find a $1.53\,$ GHz-wide (FWHM) Gaussian shape with no discernible background, see SM Fig. 7(c). (Note that the broadening compared to the VIPA specifications is due to the finite core size of the single mode collection fiber.) This yields a system resolving power $\lambda/\Delta \lambda \simeq \,1.28\,\times10^{5}$, in good agreement with the limit imposed by the VIPA. Differently stated, the cross-talk from a $5\,
$GHz detuned spectral channel is below $-25\;$dB. See SM for more information. 

Next, we create 8 spectral modes by frequency detuning the laser in steps of $6.5$ GHz (equal to the idler mode spacing, see below), and measure their spatial profiles by displacing the collection fiber. Fig. \ref{fig:spectra}(a) shows the recorded intensity variations for each of these modes. The figure also shows the expected intensity distribution, calculated by taking into account the spectral broadening imposed by the finite linewidth of the laser and the spatial broadening due to the finite size of the core of the collection fibers, %(mode field diameter$\simeq 10.4\,\mu$m)
see SM for details. The result is in good agreement with the measured data.  

After reverting to the SPDC source and the SNSPD, we assess the spatial distribution of the idler modes in the back focal plane of the lens behind the VIPA. To do this, we displace again the collection fiber, but now we record the single photon detection rates. To avoid that several spectral modes are mapped onto the same spatial position, we reduce the bandwidth of the idler photons to less than the VIPA's FSR using the $16\,$GHz-linewidth Fabry-P\'erot filter depicted in Fig.\,\ref{fig:SPDC_VIPA}. In order to capture a broad spectrum, we detune the filter cavity over a total spectral range of $130\,$GHz so that the transmission is always maximized. Using the calibration curve in SM Fig.\,7(e), 
we convert spatial positions of the fiber into frequency, and after subsequent normalization with respect to the system efficiency (SM Fig.\,7(f)), we obtain the spectrum of 20 idler modes plotted in Fig. \ref{fig:spectra}(b). As expected from the length of the cavity, we find a mode spacing of $6.5\,$GHz. Note that the total number of modes is only limited by phase matching. Estimating the total idler bandwidth to be $1.3\,$ THz, this yields 200 modes.

Next, we simulate an optical comb of identical Lorentzian peaks with the same spacing of $6.5\,$GHz, and calculate the spatial intensity distribution for different linewidth values by taking into account the response functions of the Fabry P\'erot filter and the demultiplexer (see SM, Sect. V %\ref{SM:SPDC.calc}
and Fig.\,6  %\ref{suppl.VIPAbroadenings.pdf}
). The solid blue line in Fig. \ref{fig:spectra}(b) shows the resulting spectrum for a linewidth of 1.48 GHz (FWHM). This is in good agreement with the experimentally measured spectrum and yields an approximate value of the linewidth of the idler modes produced by the source.

To complete the characterization of the cavity-enhanced SPDC emission, we also measure the detection rate of signal photons at a wavelength around $795\,$nm  as a function of detuning of the filter cavity (its transmission profile is shown in Fig.\ref{fig:spectra}(c) ).
The recorded spectrum, plotted in Fig. \ref{fig:spectra}(d), shows a mode spacing of $6.5\,$GHz, equivalent to that of the idler photons. The low contrast is due to the similarity between the mode spacing and the cavity linewidth of $6.1\,$GHz. To deduce the spectrum of the signal photons prior to filtering, we fit the measured data using the convolution of the cavity transmission profile in Fig.\ref{fig:spectra}(c) with a comb of identical Lorentzian peaks spaced by $6.5\,$GHz and varying their linewidth. We find good agreement for the expected linewidth of 1.48 GHz, calculated after taking into account the (quasi perfect) energy correlations between signal and idler photons due to the spectrally narrow pump of less than 10 MHz width. 
%\WT{which is consistent with the value for the idler photons as both spectra are linked through energy conservation}. \st{We attribute the large discrepancy with respect to the idler linewidth of around $40\,$MHz to laser frequency fluctuations during the measurement time. Note that this affects only the signal photons since the spectrum of the idler photons is determined by the (stable) SPDC cavity.}

We now assess the quantum nature of the cavity-based SPDC source. Towards this end, we first measure the 2$^{nd}$ order auto-correlation coefficients $g_{s,s}^{(2)}(0)$ and $g_{i,i}^{(2)}(0)$ for the individual signal and idler modes. 
For instance, for the signal mode, this coefficient is defined as $g_{s,s}^{(2)}(0)=p_{s,s}/p_{s}^2$, where $p_{s,s}$ is the probability of detecting two signal photons in coincidence, and $p_{s}$ is the detection probability for individual signal photons. The correlation coefficient is measured using the time tagger (see SM and \cite{askarani2020entanglement}). We obtain $g^{(2)}_{s,s}\,=\,1.299\,\pm\,0.032$ and $g_{i,i}^{(2)}=1.362\,\pm\,0.150$, indicating classical fields for which $1\leq g^{(2)}\leq 2$ \cite{zielnicki2018joint}. Due to accidental coincidences and a finite coincidence window, the auto-correlation coefficients are reduced compared to their expected value of 2 \cite{Blauensteiner2009,rielander2014}. See the supplemental materials for details. 

Second, we also measure cross-correlation coefficients $g_{s,i}^{(2)}(0)=p_{s,i}/(p_{s}\cdot p_{i})$ between signal and idler photons for various combinations of spectral channels. Here, $p_{s,i}$ is the coincidence detection probability of signal and idler photons, and $p_{s}$ and $p_{i}$ are defined above. For classical fields, the Cauchy-Schwarz parameter $R=(g_{s,i}^{(2)})^2/(g_{s,s}^{(2)}\cdot g_{i,i}^{(2)})$ is upper bounded by 1 \cite{steele2004cauchy}. Taking into account the upper limit of 2 for $g_{s,s}^{(2)}$ and $g_{i,i}^{(2)}$, we find for the cross-correlation coefficient $g_{s,i}^{(2)}\leq 2$, again under the assumption of classical fields. Our measurements are performed after replacing the fiber by the fiber array, using 5 SNSPDs instead of one, and employing the Tm$^{3+}$:LiNbO$_3\,$-based spectral filter to narrow down the spectral widths of the signal photons, as shown in Fig.\ref{fig:SPDC_VIPA} (see SM for details). Fig. \ref{fig:g2_map}(d) shows 25 cross-correlation coefficients, obtained using a laser power of $1.8\,$mW at the input of the SPDC cavity, on a grid composed of $5\times 5$ signal and idler spectral modes. The desired spectral channels were selected using a specific SNSPD (for the $1532\,$nm photons) and after tuning the Tm$^{3+}$:LiNbO$_3$ filter and adjusting the angle of the Fabry-P\'erot cavity that filter the $795\,$nm photons.  

\begin{figure}
\includegraphics[width=01\columnwidth]{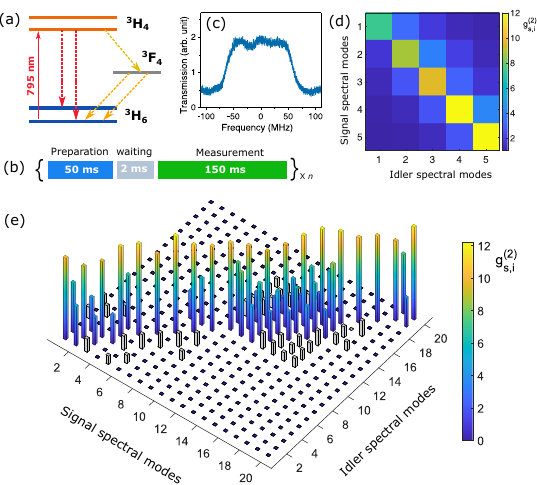}  % {fig3.pdf}
\caption{Measurements of cross-correlation coefficients. (a) Simplified Tm level scheme. (b) Measurement sequence. (c) A trench in the Tm absorption profile used to filter the signal modes. (d) 5x5 cross correlation coefficients measured with a VIPA array, 5 SNSPDs and a Tm:LiNbO$_3$ filter (e) 89 cross correlation coefficients on a 20x20 grid of signal and idler spectral modes, measured using a single collection fiber and only the $6.1\,$GHz cavity filter for the $795\,$nm photons. Panels (d) and ((e) use the same heatmap to indicate the value of g$_{s,i}^{(2)}$. 
\label{fig:g2_map}}
\end{figure}

The results of these measurements confirm our expectations. We find strong quantum correlations for energy-correlated pairs of signal/idler modes with correlation coefficients up to 10, significantly above the maximum classical value of 2. All measurements relied on a coincidence window of approximately three FWHM of the coincidence peak. A 3-fold reduction of the window width increases $g_{s,i}^{(2)}$ from 10 to 20, but at the expense of a larger uncertainty.
In addition, when either the signal or the idler mode is changed, the value for $g_{s,i}^{(2)}$ drops rapidly below 2. We attribute the small cross talk to the finite rejection of $795\,$nm photons outside the transmission bandwidth of the Tm$^{3+}$:LiNbO$_3$ filter, see Fig.\ref{fig:g2_map}(c).  

Finally, we expand the measurements to a grid of 20x20 spectral modes out of which 89 combinations were measured using the single collection fiber and no Tm$^{3+}$:LiNbO$_3$ filter. As depicted in Fig. \ref{fig:g2_map}(e) and described in the SM, we find similar results.

\emph{Discussion and conclusion.} 
Our demonstration combines the key elements of the quantum repeater architecture based on spectral multiplexing proposed in \cite{sinclair2014spectral}, namely a source of spectrally multiplexed quantum-correlated photon pairs, a compatible detection setup based on a VIPA, as well as quantum memory-anticipating spectral filtering in a Tm$^{3+}$:LiNbO$_3$ crystal. 

While the fully pigtailed, cavity-based SPDC source currently generates energy-time entangled signal and idler photons, its output can easily be adjusted to create time-bin entangled photon pairs \cite{Brendel1999} with signal and idler photons of high spectral purity, as required for multi-photon interference such as a Bell-state measurement \cite{osorio2013purity}. The width of the signal modes---currently 1.5 GHz---makes the emitted photons suitable for subsequent storage in Tm-based quantum memories using the AFC protocol \cite{saglamyurek2011broadband}, a development that we anticipate by employing the Tm-based spectral filter. (Note that the line-width can be reduced by using a cavity with higher quality factor.) However, the current mode spacing of $6.5\,$GHz should be decreased to optimize the available memory bandwidth. Furthermore, in order to increase the coincidence rate, the excess loss at $1532\,$nm wavelength needs to be removed, most likely through better pigtail alignment.

The VIPA-based detection setup allows demultiplexing of spectral modes at $1532\,$nm wavelength with $1.5\,$GHz resolution and negligible cross-talk.
To improve the setup, a VIPA with higher spectral channel density should be designed. Furthermore, to remove the coupling loss between the VIPA and the arrayed fibers, one should investigate the possibility to integrate a VIPA directly with an SNSPD array inside the cryostat. 

Taken individually, both the source and the use of the VIPA improves the state-of-the art, even before implementing the mentioned improvements. Beyond, the demonstration of both components in conjunction with the hole-burning-based spectral filter has so far been lacking. It has allowed us to reveal strong non-classical correlations between spectrally correlated signal and idler modes, and represents an important step towards a spectrally multiplexed quantum repeater based on feed-forward control.\\

\textbf{Acknowledgements.} We thank J. H. Davidson for valuable discussions, and Daniel Oblak for lending us a  795 nm Fabry-P\'erot filter cavity. We acknowledge funding through the Netherlands Organization for Scientific Research, the European Union's Horizon 2020 Research and Innovation Program under Grant Agreement No. 820445 and Project Name Quantum Internet Alliance, and the Early Research Programme of The Netherlands Organisation for Applied Scientific Research (TNO).

\bibliographystyle{apsrev4-2}
\bibliography{main}

\pagebreak

\widetext
\begin{center}
\textbf{\large Supplemental Material}
\end{center}

\maketitle

\section{A quantum repeater scheme with cavity-SPDC-based photon sources, multi-mode quantum memories and VIPA-based demultiplexer}

\begin{figure}[h]
\includegraphics[width=1\linewidth]{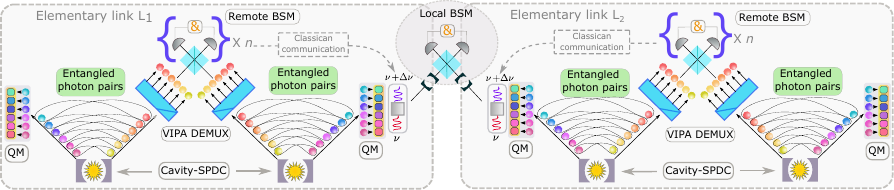}
\caption{Schematic representation of the spectrally multiplexed quantum repeater scheme proposed in \cite{sinclair2014spectral} using our cavity embedded PPLN crystal as spectrally multiplexed entangled photon sources and the VIPA set-up as demultiplexer. Frequency modes of the entangled photon pairs are represented by different colors. SPDC---spontaneous parametric down conversion.
\label{fig:Repeater}}
\end{figure}

The spectrally multiplexed quantum repeater described in \cite{sinclair2014spectral} is based on the simultaneous generation of photon pairs (composed of signal and idler photons) in many spectral modes, and a Quantum Memory (QM) with a broad absorption line that allows storing the signal photons in separate spectral channels with fixed storage time. The idler photons in each frequency mode undergo independent remote Bell State Measurements (BSM) with the photons generated by an identical source positioned on the other end of the same elementary repeater link. Entanglement swapping results in heralded entanglement between the two quantum memories of the link. Using feed-forward information (classical communication) about the spectral mode in which the BSM took place, the energy-correlated re-emitted photons from the QMs are subsequently frequency shifted to a predetermined frequency. A filter cavity allows these photons to pass, rejecting photons in all other modes. The transmitted photons from two neighboring quantum memories---one per elementary link---are then subjected to a local BSM, which swaps entanglement across the quantum communication link. SM Fig. \ref{fig:Repeater} illustrates the experimental demonstration of frequency resolved optical interface and compatibility between our cavity-embedded PPLN crystal for generating frequency-multiplexed photon pairs, a VIPA-based demultiplexing setup for performing spectrally resolved BSMs and Tm$^{3+}$:LiNbO$_{3}$ crystal for storing the signal photons. Such an efficient interface is the prerequisite for practical implementation of our repeater scheme.

%
%%%%%%%%%%%%%%%%%%% Section II. %%%%%%%%%%%%%%%%%%%%%%%%%%%%%%%%%%%

\section{Optical Design}
The main advantage of the VIPA over a prism or diffraction grating is its considerably larger angular dispersion which allows achieving high spectral resolution (10-20 times higher than a typical diffraction grating with a blaze angle of 30 degrees). Additionally, using a diffraction grating results in a polarization dependent loss, whereas dependence of the VIPA's response on the polarization of the input light is negligible \cite{shirasaki1996large, shirasaki1999virtually}.
To characterize our VIPA-based demultiplexing setup, we use a single collection fiber and translate it in x-direction to capture the light of different spectral channels. We perform frequency-resolved cross-correlation measurements first with the single collection fiber and then with an array of eight single mode fibers, five of which we use to capture five spectral modes simultaneoulsy. %fiber will be replaced by a fiber array to capture the light from all spectral modes simultaneously. 
In this section we describe the setup based on the single collection fiber; the setup with the fiber array is described in section IX. 

\begin{figure}[h]
\includegraphics[width=0.8\linewidth]{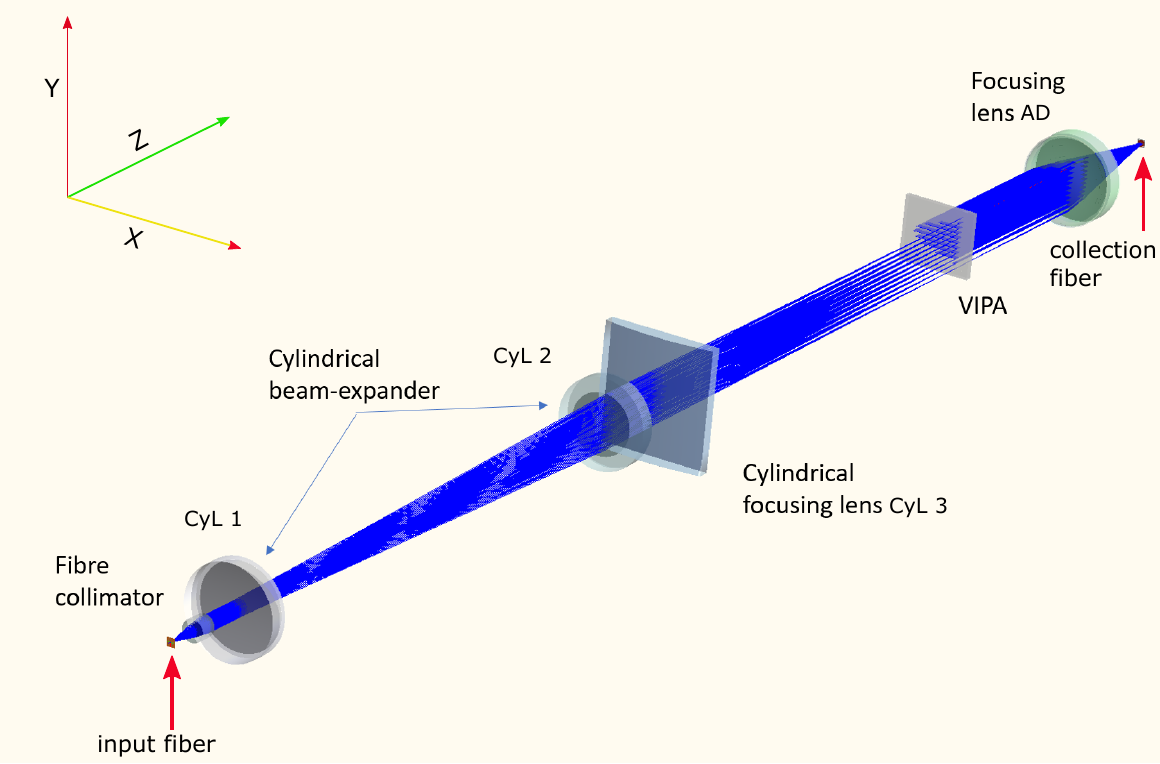}
\caption{Optical layout of the VIPA setup, shown from the fiber output at the source side (left), up to the input to the collection fiber at the detection side (right).
\label{fig:Optics}}
\end{figure}

A schematic of the optical system of the frequency demultiplexer is depicted in SM Fig. \ref{fig:Optics}. 
At the exit of the input optical fiber that delivers the multiplexed photons, we place a fiber collimator that produces a circular cross-section beam with a diameter of about $2.1\,$mm.  For wavelengths around $\lambda _0\,=\,1532.71\,$nm, the width of the outgoing VIPA beam in the horizontal (x) direction is $\sim 8\,$mm (value specified by the manufacturer and determined by the number of reflections inside the VIPA). To obtain a circularly symmetric beam, as required for optimal coupling of the light into the collection fiber placed behind the VIPA, the beam from the input fiber is expanded 4$\times$ in the vertical (y) direction. Towards this end, we used a cylindrical beam expander. It consists of a negative $f=-50\,$mm  and a positive $f=200\,$mm cylindrical lenses, where $f$ denotes the focal length. After the beam expander, we place a cylindrical lens of $f=150\,$mm focal length, which creates a line-shaped focal spot at the entrance slit of the VIPA.

%A schematic of the optical system of the frequency demultiplexer is depicted in SM Fig. \ref{fig:Optics}. At the output of the input optical fiber that delivers the multiplexed photons, we place a fiber collimator that produces a circular cross-section beam with a diameter of about $2.1\,$mm. To achieve optimal light coupling into the collection fiber behind the VIPA, a beam with a circular cross section is required. Since for 9 spectral channels around the selected wavelength the VIPA unit produces a $\sim 8\,$mm long array of light spots in the x-direction, the beam from the input fiber is 4$\times$ expanded in the y-direction. To achieve this, we used a cylindrical beam expander. It consists of a negative $f=-50\,$mm  and a positive $f=200\,$mm cylindrical lenses, where $f$ denotes the focal length. After the beam expander, we place a cylindrical lens of $f=150\,$mm focal length, which creates a line-shaped focal spot at the entrance slit of the VIPA.

At the output of the VIPA, individual spectral modes propagate as parallel beams with about equal dimensions in x and y directions, and each of the modes exits the VIPA at a slightly different angle. To couple the light from a given spectral mode optimally into a SMF-28 single mode fiber, 
we use a spherical achromatic doublet lens (AD) with focal distance $f\, = 45\,$mm. This lens maps the exiting spectral modes---all occupying different spatial modes---onto different spots in its back-focal plane. The collection fiber is mounted on a translation stage with five degrees of freedom (three translational and two rotational ones). 

%

%%%%%%%%%%%%%%%%%% Section III. %%%%%%%%%%%%%%%%%%%%%%%%%%%%%

\section{Analytical modelling of the VIPA output in the back focal plane of the collection lens.}
To describe the wavelength-dependent field distribution at the output of the VIPA setup, we use the model put forward by S. Xiao et al. \cite{xiao2005experimental}. In the y-direction, the beam shape of each peak in the back focal plane of the focusing lens AD is described by a  Gaussian, with a FWHM defined by the focal length of the AD and the effective focal length of the cylindrical beam expander (cylindrical lenses CyL1 with focal distance of f$_{CyL1}=-50\,$mm and CyL2 with focal distance f$_{CyL2}=200\,$mm). Therefore, we will mainly focus on the VIPA output in the spectral demultiplexing direction, x.
In our setup, the field distribution along the x-direction (at $y=0$) in the back focal plane of the out-coupling lens AD, which focuses the output beam into the collection fiber, is described by:

\begin{equation}
\label{eq:vipafield}
E_{out}(x,\lambda) \propto  e^{-\frac{f_1^2 x^2}{f_2^2 W^2}} \bigg[ 1-r R \exp \bigg( \frac{2i\pi t}{\lambda} \Big( \frac{x^2 \cos(\theta_{in} n)}{f_2^2\;n} +\frac{2\;x\;\tan(\theta_{in})\;\cos(\theta_{in}\;n)}{f_2}\;-\;2 \;n\;\cos(\theta_{in}) \Big) \bigg) \bigg] ^{-1}.
\end{equation}

Here, $n$ is the refractive index of the VIPA material, $t$ is the VIPA thickness, $R$ and $r$ are the reflectivities of the VIPA's front and back surfaces, respectively, $\theta_{in}$ is the incidence angle of the incoming light beam onto VIPA, $W$ is the radius (in x-direction) of the collimated input beam (i.e. before the lens focusing it onto the VIPA slit), $f_1$ is the focal length of the cylindrical lens CyL3 focusing the beam onto the entrance slit of the VIPA in x-direction, and  $f_2$ is the focal length of the spherical lens AD focusing the VIPA output beam into the optical fiber on the collection/detection side.

\begin{figure}%[h]
%\centering
\begin{subfigure}[b]{0.5\linewidth}
    %\centering
    \includegraphics[width=1\linewidth]{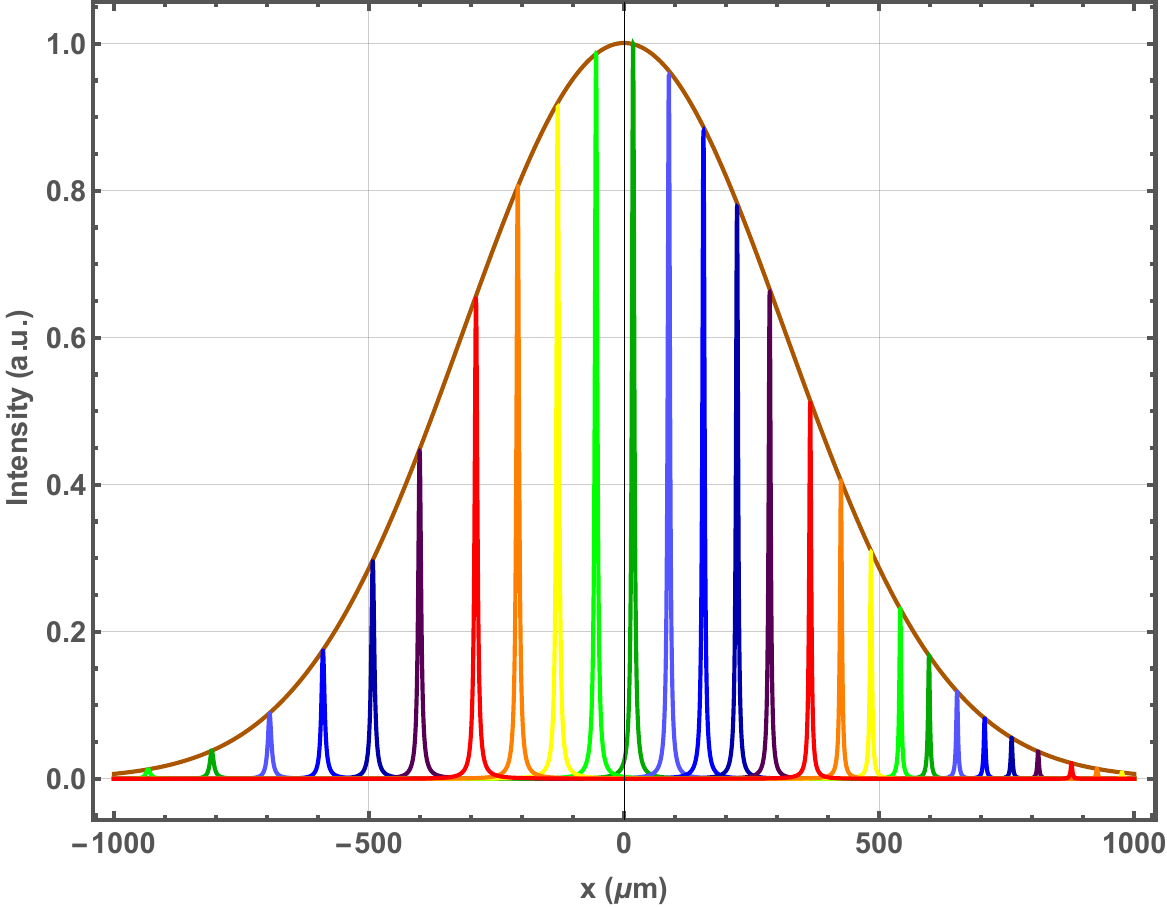}  
    \caption{}
    \label{VIPA-9freqs}
    \end{subfigure}
    \hfill
%\end{figure*}
%
%\begin{figure*}[h]
    \begin{subfigure}[b]{0.48\linewidth}
    \includegraphics[width=1\linewidth]{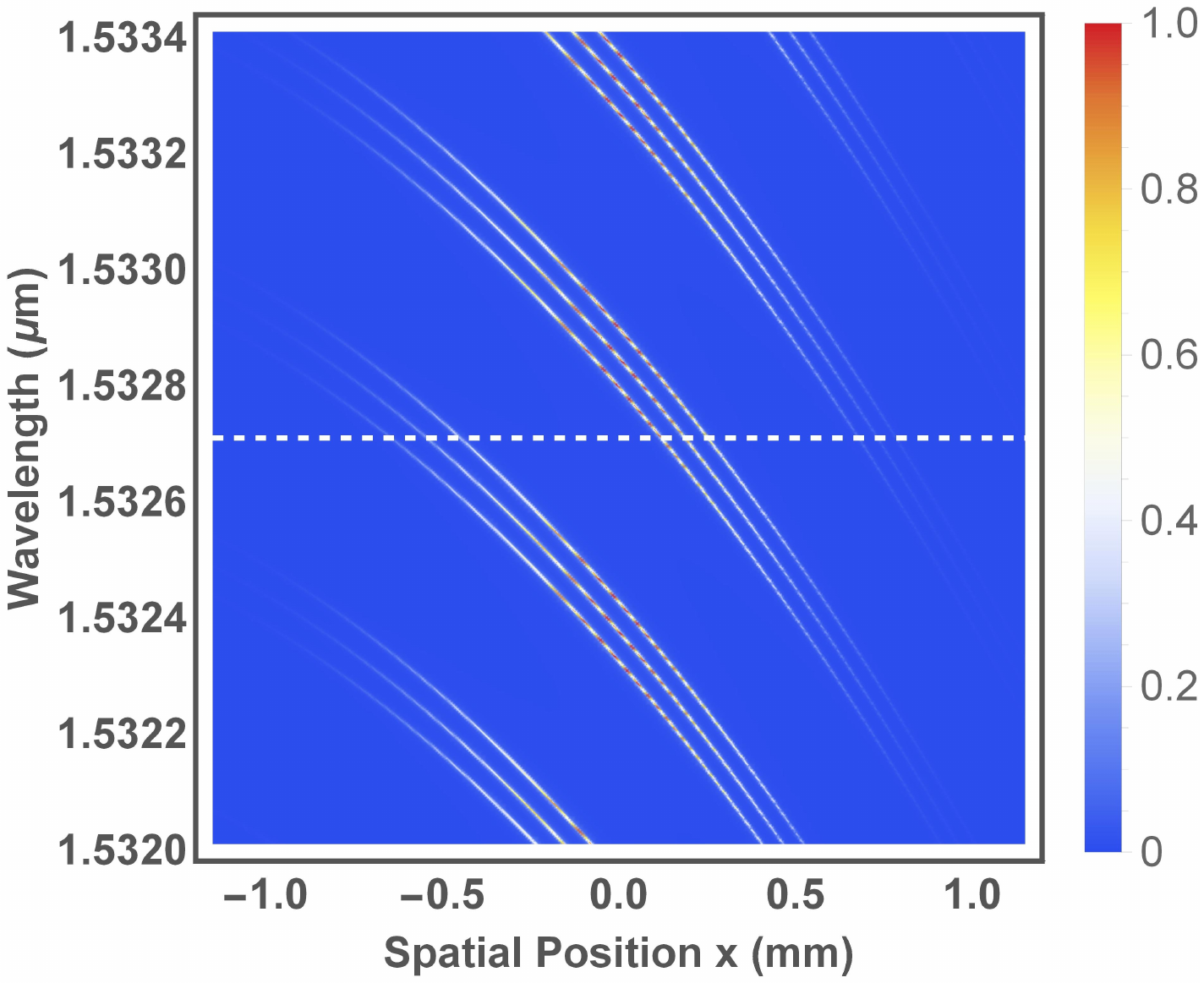}
    \caption{}
    \label{VIPAint_x-and-wavel}
    \end{subfigure}
    \hfill
\label{fig:VIPA.calculs}
\caption{(a) Brown line: envelope of the intensity of different spectral modes at the input of the collection fiber. Colored lines of other colors  (rainbow-like, from red to dark magenta): calculated VIPA response for 9 different single-frequency channels at around $\lambda_0 =1532.71\;$nm, spaced by $6.5\,$GHz. The highest-intensity-order peaks of the nine frequency channels are situated in the interval between spatial coordinates $x\,=-\,450\,\mu$m and $x=\,330\,\mu$m. Neighbouring diffraction orders are situated at smaller and larger values of the x-coordinate. (b) Intensity at the input of the collection fiber as a function of wavelength and spatial position measured along the x-axis, plotted for the central spectral channel and for the two neighboring spectral channels ($\lambda \pm\,6.5\,$GHz). The dashed line shows the wavelength of the central spectral channel in our experiment.}
\end{figure}

The spectrum of the VIPA response predicted by this model for our setup, is plotted in SM Fig. \ref{VIPA-9freqs}. The brightest order, as well as the two neighbouring orders, are visible. We find that the maximum number of channels with $6.5\,$GHz separation that can be used simultaneously for this particular VIPA without mapping different spectral channels onto the same spatial position and without significant increase in cross-talk, is limited to about 9.

The dependence of the VIPA device's output intensity on both wavelength and spatial coordinate x (in the demultiplexing direction) is plotted in SM Fig. \ref{VIPAint_x-and-wavel} for the central wavelength $\lambda_0 = 1532.71\,$nm and the two neighbouring spectral channels separated from the central one by $\pm \, 6.5\,$GHz. The figure clearly shows different VIPA orders in the x-$\lambda$ plane (along a constant-wavelength line). We see that both the spatial positions of intensity maxima and the distances between the spatial positions of the neighbouring spectral channels vary in a non-linear fashion with wavelength. Therefore, the optimal position of the collection fiber (or the distances between collection fibers in a fiber array) are a non-linear function of the wavelength, as mentioned in the main article on p.3. Another insight provided by this figure is that it may be possible to further reduce the cross-talk between the neighbouring spectral channels (separated by $6.5\,$GHz) by shifting the operation wavelength by one or a few FSRs to shorter wavelength. However, further optimization of the cross-talk between spectral channels is outside the scope of this work.

%%%%%%%%%%%%%%%%%% Section IV. %%%%%%%%%%%%%%%%%%%%%%%%%%%

\section{Modelling the spectral and spatial characteristics of the VIPA setup.}

In the following we describe how we calculated the theoretical curves that are shown in Fig. 2(a) of the main text.

The E-field of the mode of a single-mode optical fiber along its diameter in x-direction is described by a Gaussian:
\begin{equation}
\label{eq:ESMF28}
E_{SMF28}(x) = E_0 e^{\frac{-(x-x_0)^2}{a^2}}
\end{equation}
Here, $E_0$ is the maximum amplitude of the input E-field, $x_0$ is the position of the fiber core's center, and $a$ is the mode field radius of the fiber mode. For the standard SMF-28 fiber used here, $a = 5.2\,\mu$m (corresponding to a mode field diameter of $10.4\,\mu$m). In a similar way, one can describe the E-field distribution in y-direction.

In our setup, the beam exits the VIPA, passes through the focusing lens AD (f = 45 mm), and is then coupled into an optical fiber that is placed in the back-focal plane of this lens. Therefore, due to the finite size of the collection fiber's core, for a perfectly monochromatic input E-field (i.e. a field with a delta-like frequency spectrum), the signal intensity at the output of the optical fiber is described by the convolution of the E-field at the VIPA output with the E-field of the fundamental mode of the optical fiber (we assume no transmission loss):
\begin{equation}
\label{eq:fieldsconv}
E_{out}(x, \lambda) = \int_{-\infty}^{+\infty} {E_{x, SMF28}(\tau-x) \,E_{x, VIPA}(x, \lambda) d\tau}
\end{equation}

To calculate the associated intensity profile for each spectral channel (the integration over which is proportional to the measured signal), we evaluate $I_{out}(x, \lambda) = |E_{out}(x, \lambda)|^2$, with integration boundaries in eq.(\ref{eq:fieldsconv}) replaced by the spatial values fully enclosing the associated spatial order of the VIPA peak.

However, in reality the input field is not monochromatic but is described by a Lorentzian wavelength distribution: 
\begin{equation}
    f(\lambda) = \left(1+\, \frac{(\lambda - \lambda_0)^2}{b^2} \right) ^{-1},
\end{equation}
where \textit{b} is the half-width at half-maximum of the distribution, and $\lambda_0$ is its central wavelength.

Therefore, to model a realistic signal similar to the one obtained in our experiments, we calculate the sum of fiber-coupled VIPA responses at each wavelength $\lambda_i$ in the input spectrum, multiplied by its respective weight (i.e. its relative intensity in the spectrum):
\begin{equation}
\label{eq:spectralbroadening}
I_{out, tot}(x)= \sum_{i}^{}{f(\lambda_i) I_{out}(x, \lambda_i)}
\end{equation}

\begin{figure}[h]
\includegraphics[width=0.7\linewidth]{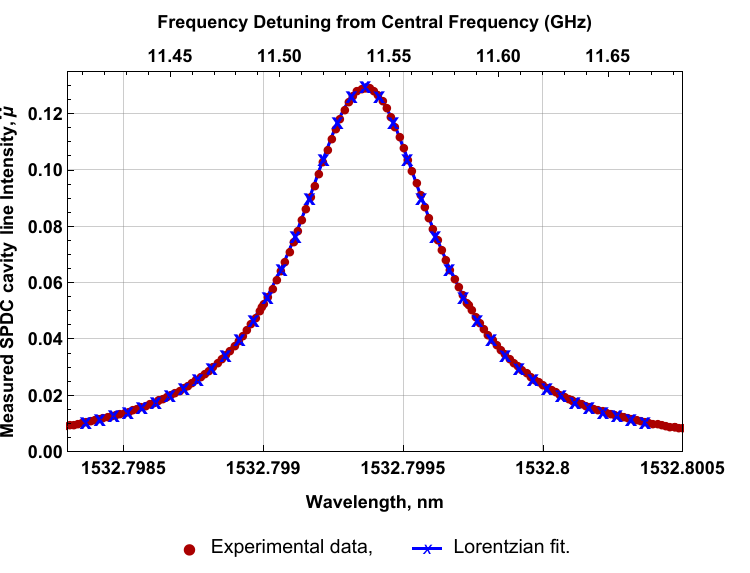}
\caption{Spectral shape of the input laser line used for spectral characterization of VIPA setup. Red dots: experimental data, blue crosses and line: Lorentzian fit to the data.
\label{fig:cavityline}}
\end{figure}

The spectrum of a narrow-line tunable laser, used as input light, was measured experimentally, and fitted with a Lorentzian. Both the measured data and the fit are shown in SM Fig. \ref{fig:cavityline}.

\begin{figure}[h]
\includegraphics[width=0.8\linewidth]{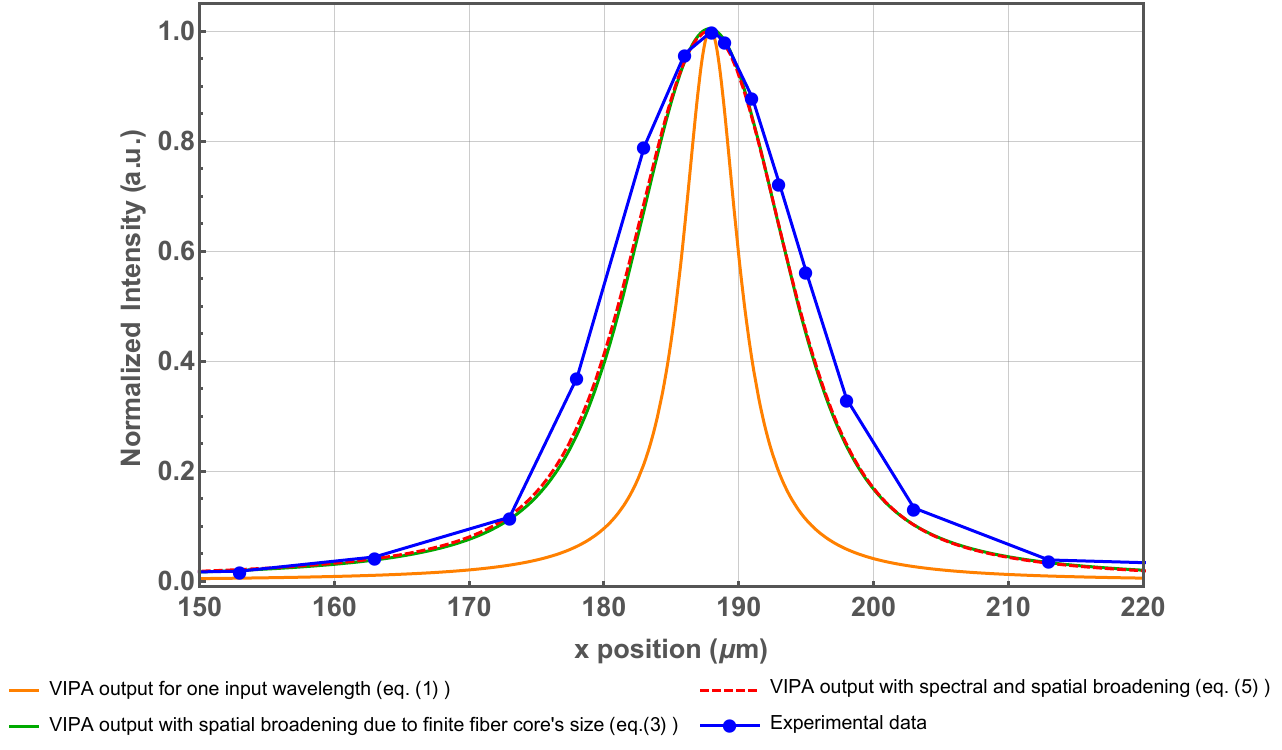}
\caption{Effects of the finite size of the fiber core and of the spectral broadening onto the calculated VIPA line shape, and comparison of the result with experimental data.
\label{fig:vipa-broadening}}
\end{figure}

This allows us to calculate the VIPA setup response, i.e. the signal measured atthe output of the collection fiber as a function of the fiber's position. As an example, in SM Fig.\ref{fig:vipa-broadening} we plot the calculated response for each step as described by equations (\ref{eq:vipafield}-\ref{eq:fieldsconv}, \ref{eq:spectralbroadening}), along with the experimental data, for a central laser wavelength of $\lambda=\,1532.71\,$nm. Small deviations of experimental data from the calculations are likely due to deviations from the paraxial approximation model and imperfections in setup alignment. The obtained result indicates that spatial broadening of the VIPA's line is mainly due to the finite core size of the collection fiber, while the effect of the finite linewidth of the laser is very small.

%%%%%%%%%%%%%%%%% Section V. %%%%%%%%%%%%%%%%%%%%%%%%%%

\section{Deducing the SPDC spectrum from measured data.}
\label{SM:SPDC.calc}

In order to deduce the linewidth of the idler modes at around 1532 nm wavelength in the SPDC spectrum, we use the data recorded with SPDC light (see Fig. 2(b) in the main text). They contain the normalized photon detection rates as a function of both laser frequency detuning and of fiber position along the x direction. The data were acquired as described in the main text. 

We take the following steps. 

\begin{figure}[h]
\includegraphics[width=0.6\linewidth]{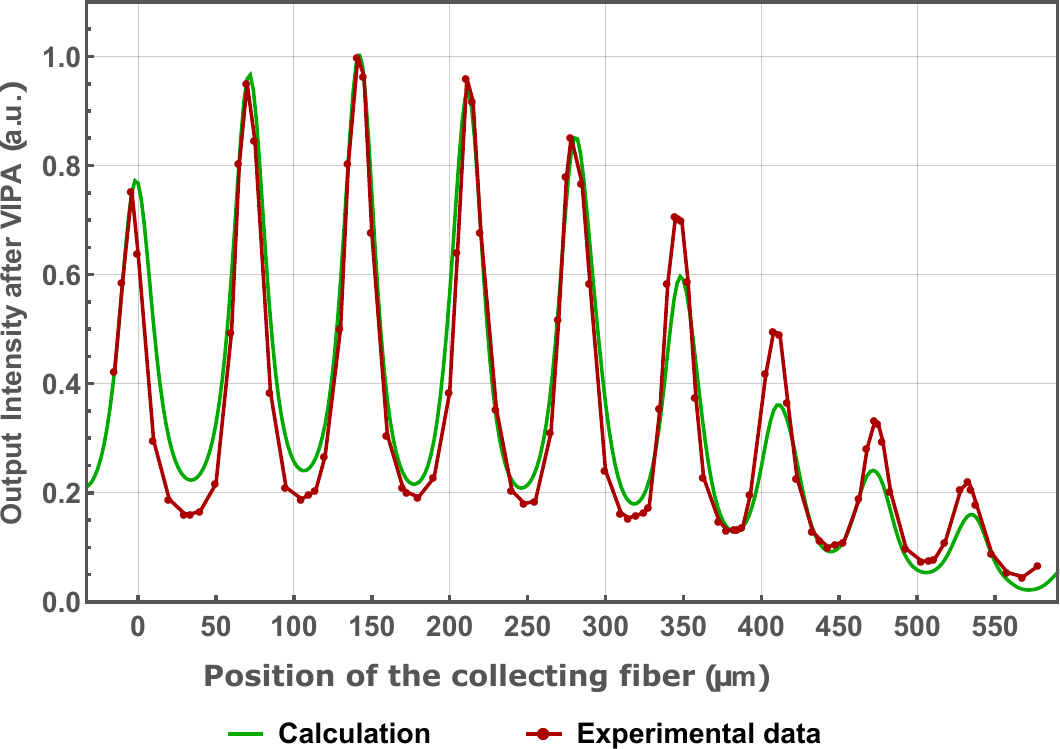}
\caption{Normalized spectrum of the VIPA output. Green: calculated as described above, with Lorentzian linewidth of  1.48\,3stGHz. Red: experimental data, measured by translating the collection fiber horizontally in the focal plane of the collection lens. %as described in the main text of the article.
\label{fig:SPDCfit}}
\end{figure}

\begin{enumerate}
    \item To simulate the SPDC spectrum, we define an $E$-field spectrum consisting of a sum of a finite number of identical Lorentzians with central wavelengths spaced by about $6.5\,$GHz. The width of the Lorentzians is the only parameter that is tuned in order to achieve a good overlap with the experimentally measured data.
    \item We adjust the central wavelengths of these Lorentzians to position them at the same wavelengths as in the experimentally measured data set (Fig. 2(b) of the main text). 
    \item  We calculate the effect of the Fabry-P\'erot cavity used for filtering (FWHM $\sim\,16\,$GHz) onto the E-field spectrum. This allows us to include the spectral filtering effects into the model/calculation.  
    \item We calculate the convolution of the total E-field obtained in step (3) with the $E$-field of the VIPA calculated by equation (\ref{eq:vipafield}). This gives the total $E$-field at the VIPA output.
    \item Then we take into account the spatial broadening due to finite core size of the SMF-28 optical fiber. We do this by evaluating the convolution of the total E-field at the VIPA output (calculated in step 4) with the $E$-field of the mode of the SMF-28 fiber, as described by equation (\ref{eq:fieldsconv}).
    \item Next, we evaluate the intensity of the resulting $E$-field by taking its amplitude-square value. The resulting quantity is proportional to the intensity of the $E$-field measured through the collection fiber (times a constant).
    \item Finally, we normalize the resulting spectrum and compare it to the (normalized) experimental data. We continue adjusting the linewidth of the Lorentzian lines of the initial SPDC spectrum until a good qualitative agreement with the linewidths of the measured idler mode peaks is achieved (see SM Fig. \ref{fig:SPDCfit}). 
\end{enumerate}

The obtained normalized intensity spectrum of the $E$-field of step (1) with Lorentzian's width determined in step (7), is plotted in Fig. 2(b) of the main text as a solid line. The linewidth of the Lorentzians in this frequency comb gives a good estimate of the linewidth of the SPDC modes. An example of the resulting calculation for one spectral position of the Fabry-P\'erot filter cavity (see description of data collection for Fig. 2(b) in the main text) is plotted in SM Fig. \ref{fig:SPDCfit} (green line) along with the experimental data (red).
For the Lorentzian linewidth of about 1.48\,GHz, the peak widths in the calculated VIPA spectrum (green line in SM Fig. \ref{fig:SPDCfit}) are in good qualitative agreement with the experimentally measured data (red line and dots in SM Fig.\ref{fig:SPDCfit}). Hence, we estimate the approximate linewidth of the idler modes of our SPDC source to be close to 1.48\,GHz.  

%%%%%%%%%%%%%%%%%%% Section VI. %%%%%%%%%%%%%%%%%%%%%%%%%%%%

\section{Mapping spectral modes to spatial channels: cross-talk and calibration data}

\begin{figure}[h]
\includegraphics[width=0.8\linewidth]{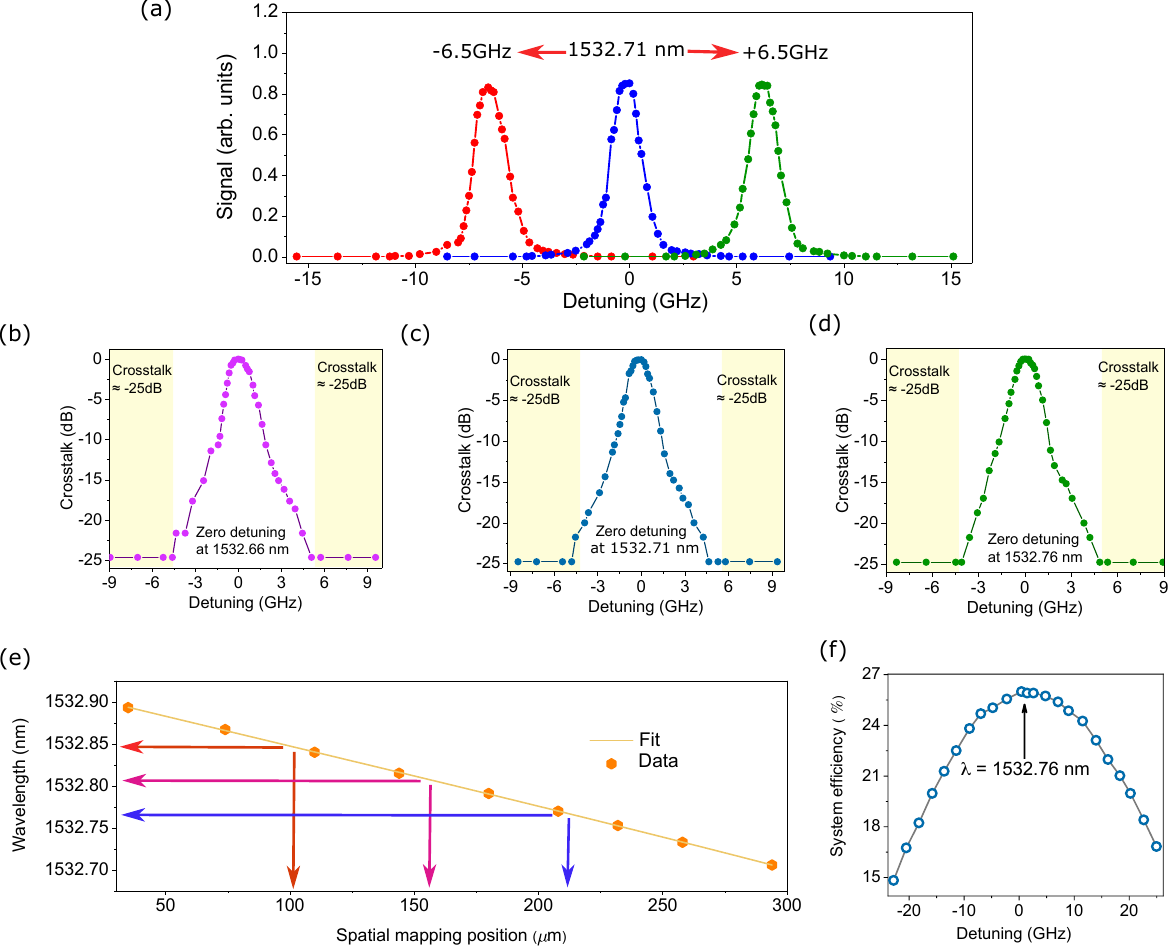}
\caption{(a) Measured signal in the collection fiber as a function of frequency detuning for light at $1532.71\,$nm wavelength and light detuned by $\pm{6.5}$ GHz to $1532.66\,$nm and $1532.76\,$nm. Variation of cross-talk for the spectral channels at (b) $1532.66\,$nm and (c) $1532.76\,$nm  (d) Calibration curve for mapping spectral to spatial modes. 
\label{fig:VIPA_charctrzn}}
\end{figure}

To ensure that we collect the spectral modes with $6.5\,$GHz spacing consistently with very low cross-talk, we investigate the cross-talk for two other channels, in addition to the spatial channel corresponding to $1532.71\,$nm wavelength discussed in the main text. We couple the signal by positioning the SMF core at the mapping positions for $1532.66\,$nm and $1532.76\,$nm wavelengths (detuned by $\pm\,6.5\,$GHz from $1532.71\,$nm) and measure the signal with frequency detuning of the laser. The results are shown in SM Fig. \ref{fig:VIPA_charctrzn}(a). From these data we calculate cross-talk as a function of frequency detuning for these two channels, shown in SM Fig.\ref{fig:VIPA_charctrzn}(b) and (d). We find the cross-talk to be below $-25\,$dB (limited by the resolution of the measurement setup) for a detuning of more than $\pm\,5\,$GHz for both channels.

We detune the laser's frequency and record the spatial mapping positions of the spectral lines in the back-focal plane of the achromatic doublet (AD) behind the VIPA. Theses results allow us to obtain the calibration curve $\lambda(x)$ for mapping spectral modes to spatial channels. Here, $x$ is the mapping position in $\mu$m. For the spectral lines ranging from $1532.9127$ to $1532.4374\,$nm, we fit the calibration data using the polynomial 

\begin{equation}
\label{eq:calbrn}
\lambda(x) = 1532.9127-(7.03138\times10^{-4})\,x-(6.7756\times10^{-8})\,x^2-(6.3004\times10^{-11})\,x^3,
\end{equation}

\noindent
which is depicted in SM Fig.\ref{fig:VIPA_charctrzn}(e). Using this calibration, we generate the idler spectrum shown in Fig.2(b) in the main text. 

SM Fig. \ref{fig:VIPA_charctrzn}(f) shows the variation of the system efficiency of the demultiplexer as a function of wavelength, measured again using a laser and a linear photo-detector and after optimizing the position of the collection fiber for each wavelength. The system efficiency shows a Gaussian dependence with a maximum value of $25.5 \%$, which drops down to about $17\%$ as the frequency is detuned by $-20$ or $+25\;$GHz. Note that the optimal position of the collection fiber changes in a non-linear fashion with the wavelength (see SM Fig. \ref{VIPA-9freqs}).

%%%%%%%%%%%%%%%%%%%%%%%%%%%%%%%%%%%%%%%%%%%%%%%%%%%%%%%%%%%%%%%%

\section{Auto-correlation coefficients}

\begin{figure}[h]
\includegraphics[width=0.7\linewidth]{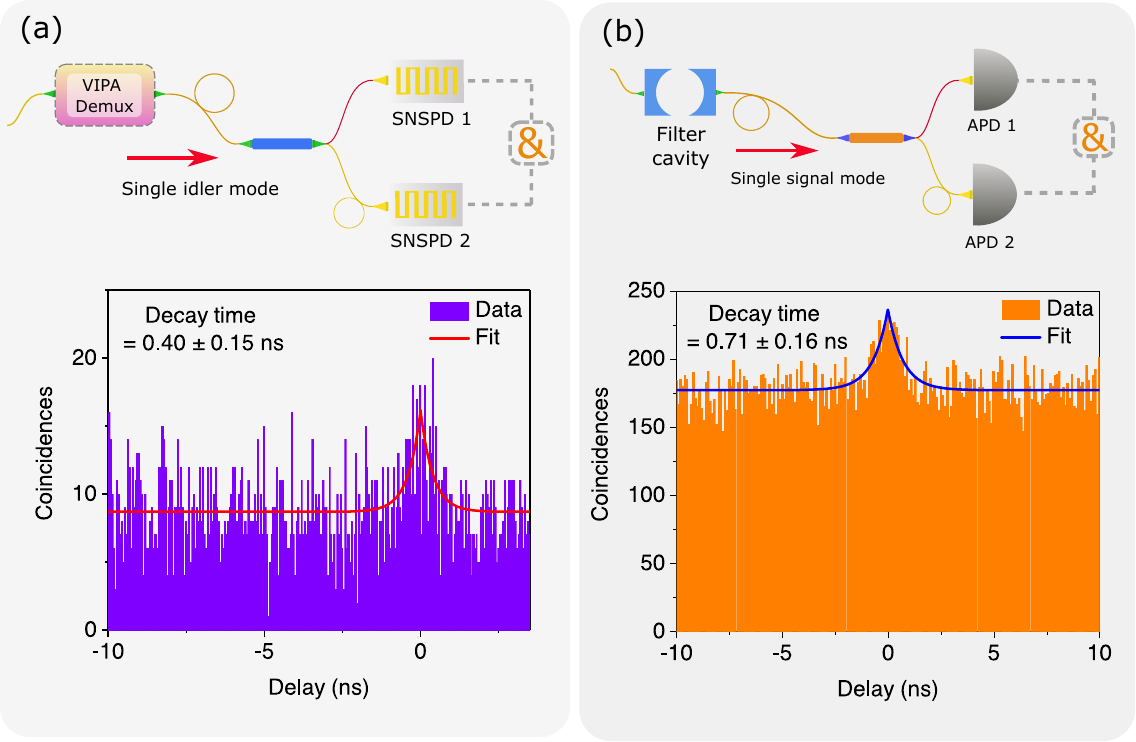}
\caption{Time-resolved coincidence-count histograms collected using a single spectral mode of the (a) idler and (b) signal 
photons, selected using the VIPA-based demultiplexer (at 1532 nm) and the filter cavity (at 795 nm). Measurement times were 20 and 12 hours, respectively. The data are fitted to symmetric exponential functions that are shown by the solid lines. The insets illustrate the HBT interferometers. 
\label{fig:autocorn}}
\end{figure}

We measure the second order auto-correlation coefficients $g^{(2)}_{s,s}$ and $g^{(2)}_{i,i}$ for individual spectral modes of the signal and idler photons using Hanbury-Brown and Twiss (HBT) interferometers. After spectrally selecting a single idler mode using the VIPA-based demultiplexer, we split the output signal using a 50:50 beam splitter, and measure the time-resolved coincidence count histogram using two Superconducting Nanowire Single-Photon Detectors (SNSPDs) and a time tagger. SM Fig. \ref{fig:autocorn}(a) shows the histogram with the HBT setup in the inset. We integrate the coincidences contained in the peak over a temporal window of $\delta t\,=\,1.4\,$ns. Dividing this number by the number of accidental coincidences, integrated over another $1.4\,$ns wide window next to the coincidence peak, yields $g_{i,i}^{(2)}\,=\,1.362\,\pm\,0.150$. In a similar way, we spectrally filter a signal mode using the Fabry-P\'erot cavity FP$_{signal}$ of $6.1\,$GHz bandwidth (FWHM), and an additional cavity filter of of 1.5 GHz bandwidth (again FWHM), which makes sure that no spurious signal mode is transmitted. We measure the time-resolved coincidence histogram that is shown in Fig. \ref{fig:autocorn}(b). In this case the coincidence window is  $1.2\,$ns wide, and we obtain $g^{(2)}_{s,s}\,=\,1.299\,\pm\,0.032$. We perform all these measurements at a pump power of $1\,$mW.

As described in \cite{Blauensteiner2009,rielander2014quantum}, the expected value for the auto-correlation coefficient is 2 if noise is negligible and if the coincidence time window $\delta t$ of the detection unit is small compared to the correlation time $\tau_c$ of the photons (which manifests itself in an auto-correlation peak of finite width). However, these conditions are not satisfied in our case. Indeed, for the idler photons, the small ratio between photon detection rates and dark count rates reduces the measured auto-correlation function. Using photon detection rates 
$S_A$=280 Hz and $S_B$=280 Hz, dark count rates $D_A$=70 Hz and $D_B$=120 Hz, and total rates $N_A$=$S_A$+$D_A$=350 Hz and $N_B$=$S_B$+$D_B$=400 Hz (where the indices $A$ and $B$ denote the two different detectors), we find \cite{rielander2014quantum}
\[
g^{(2)}(0)^{theo}_{i,i}\,=1+\frac{S_AS_B}{N_AN_B}=1.56.
\]
A further reduction is due to the use of a coincidence window of width $\delta t$. From Eq. 6 in the Supplemental Material of \cite{rielander2014quantum} (after correcting a typo), and using $\delta t = 1.4\,$ns and $\tau_c = (400 \pm 150)\,$ps (obtained by fitting symmetric exponential decays to the auto-correlation peak in Fig. \ref{fig:autocorn}(b)), we obtain
\[
g^{(2)}(\delta t)^{theo}_{ii} = (g^{(2)}(0)^{theo}_{ii}-1)\frac{2\tau_c}{\delta t}\big(1-\exp{(-\frac{\delta t}{2\tau_c})}\big)+1 = 1.26 \pm 0.06.
\]
This is consistent with the experimentally measured value $g_{i,i}^{(2)}$=1.362$\pm$0.150.
 
Similarly, we find for the signal photons 
\[
g^{(2)}(0)^{theo}_{s,s}\,= 1.83.
\]
Here we used $S'_A$=1600 Hz, $S'_B$=1600 Hz, $D'_A$=70 Hz, $D'_B$=250 Hz, $N'_A$=$S'_A$+$B'_A$=1670 Hz and $N'_B$=$S'_B$+$B'_B$=1850 Hz. Finally, $\delta t'$=1.2 ns and $\tau'_c = (710 \pm 160)\,$ps yields
\[
g^{(2)}(\delta t)^{theo}_{ss} = 1.56 \pm 0.06
\]
which is again not far from the measured value $g^{(2)}_{s,s}\,=\,1.299\,\pm\,0.032$. %The remaining difference, in particular the reason for the overestimated value of $g^{(2)}(0)^{theo}_{s,s}$, is still a matter of investigation. 

%%%%%%%%%%%%%%%%%%%%%%%%%%%%%%%%%%%%%%%%%%%%%%%%%%%%%%%%%%%%%%%%%%%%%%%
%
\section{Second order cross-correlations}

To identify and select energy-correlated frequency modes for signal and idler photons, we optimize the count rate of the $795\,$nm photons by tuning the FP$_{signal}$ cavity to $795.325\,$nm wavelength. As the signal modes are spaced by $6.5\,$GHz, the FP$_{signal}$ cavity with line-width LW$=\,6.1\,$GHz allows to select predominantly a single spectral mode (the two neighboring modes are also transmitted, but with significant loss). 
We maximize the coincidence count rate by tuning the FP$_{idler}$ cavity, first without the VIPA-based demultiplexer. Since the $16\,$GHz-large bandwidth of this cavity is much broader than the idler mode spacing of 6.5 GHz, we subsequently add the VIPA-based demultiplexer to select individual idler modes with negligibly small cross-talk from neighbouring modes. This allows us to identify the energy-correlated signal/idler pair of modes, i.e. the modes for which the cross correlation coefficient is at its maximum. 

\begin{figure}
\includegraphics[width=0.7\linewidth]{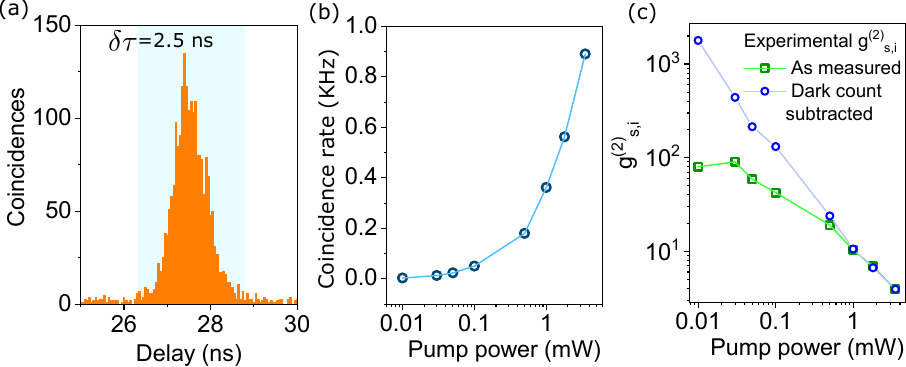}\caption{(a) Time-resolved coincidence-count histogram between signal and idler photons for pump power of 0.5 mW. Spectral modes are filtered using the Fabry-P\'erot cavity and the frequency-resolved VIPA setup for the signal and idler photons, respectively. (b) Coincidence count rates and (c) cross correlation coefficients $g_{s,i}^{(2)}$ for  spectrally correlated signal and idler modes as a function of pump power without and with dark counts subtracted. \label{fig:coincidences}}
\end{figure}

SM Fig. \ref{fig:coincidences}(a) depicts a time-resolved coincidence-count histogram using such a pair of correlated signal-idler spectral modes. Coincidences outside the peak are due to uncorrelated photons and detector dark counts. We obtain the total coincidence counts by integrating the counts within the correlation interval ($\delta\tau=2.5\,$ns), and calculate $g_{s,i}^{(2)}$ by dividing the integrated coincidences by the noise counts accumulated over a time window of the same width \cite{Grimau2020}. The coincidence count rate within the detection window and $g_{s,i}^{(2)}$ are plotted in SM Fig. \ref{fig:coincidences}(b) and (c) as a function of the pump power. The reduction of $g_{s,i}^{(2)}$ with increasing power indicates the increase in multi-photon emissions. For pump power $<\,50\,\mu$W, the detector dark counts dominate the single photon detection events, leading to a reduction in $g_{s,i}^{(2)}$. For a pump power of $10\mu$W, the as-measured and the dark-count-subtracted $g_{s,i}^{(2)}$ values are $80$ and $1784$, respectively. These values are well above the classical limit of 2, demonstrating strong non-classical correlation between the signal and idler modes.

\begin{figure}
\includegraphics[width=0.55\linewidth]{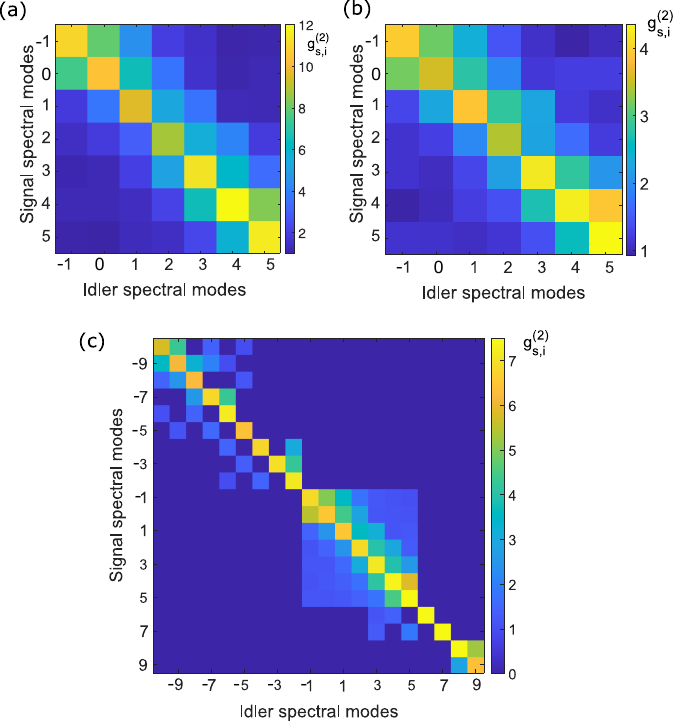}
\caption{Map of cross-correlation coefficients between $7 \times 7$ spectral modes of the signal and idler photons for a pump power of (a) 1 mW and (b) $3.5\,$mW. (c) Map of a subset of $20\times 20$ spectral modes for a pump power of $1.8\,$mW.
\label{fig:JSI_1p8mW_3p5mW}}
\end{figure}

To move to the neighbouring signal mode, we detune the FP$_{signal}$ cavity by $6.5\,$GHz. To find the corresponding energy-correlated idler mode, we shift the collection fiber at the VIPA output until we have again maximized the coincidence detection rate. This yields the position onto which the neighbouring idler mode, which satisfies energy conservation, is mapped. Repeating this procedure, we can establish correlation coefficients between any pair of signal and idler modes.

SM Fig. \ref{fig:JSI_1p8mW_3p5mW}(a) and (b) depict the correlation coefficients for a grid of $7 \times 7$ signal and idler frequency modes measured using pump powers of $1\,$mw and $3.5\,$mW, respectively. One can observe that $g^{(2)}_{s,i}>\,2$ for all energy-correlated modes, signifying non-classical correlations, and a reduction of $g^{(2)}_{s,i}$ to the classical limit for uncorrelated modes. Note that at a pump power of $3.5\,$mW it is possible to simultaneously obtain non-classical correlations (with $g_{s,i}^{(2)}\simeq 4$) and a (comparatively) high coincidence rate of $\simeq0.9\,$kHz (signal count-rate: $708.874\,$kHz, idler count-rate: $2.323\,$kHz).

\begin{figure}
\includegraphics[width=0.7\linewidth]{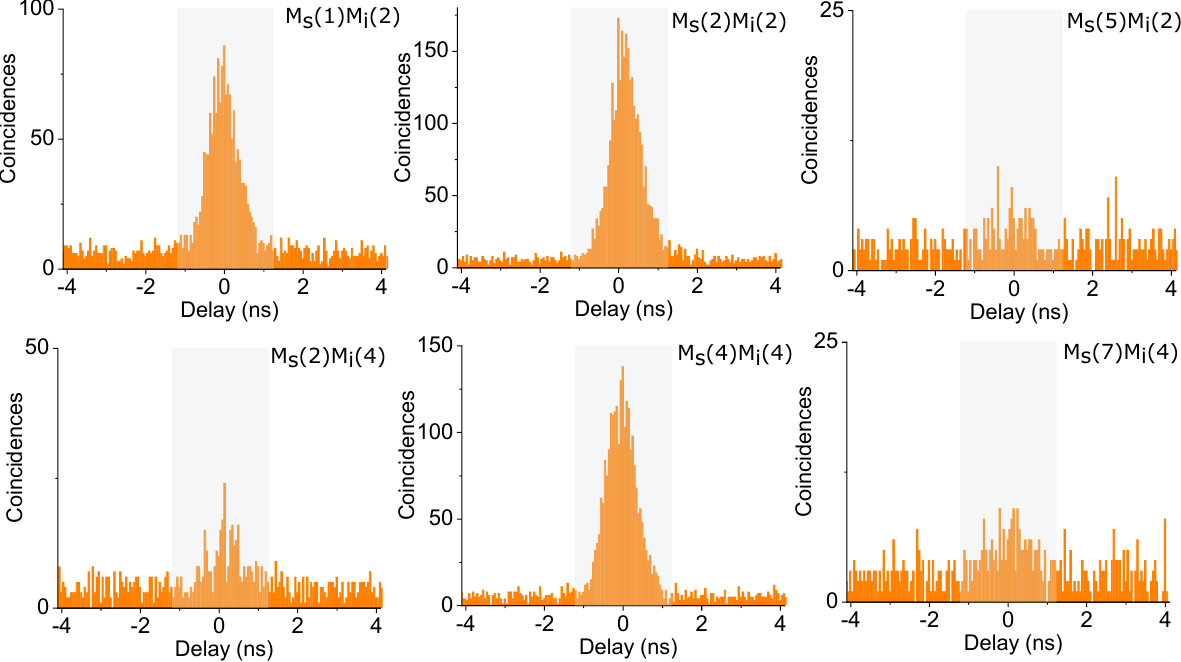}\caption{Time-resolved coincidence histograms for certain energy-correlated and energy-uncorrelated modes. Grey shading identifies the correlation intervals. \label{fig:histograms}}
\end{figure}

SM Fig.\ref{fig:JSI_1p8mW_3p5mW}(c) shows 89 $g^{(2)}_{s,i}$ values taken with a pump power of $1.8\,$mW (the plot in (a) depicts a subset). They are arranged on a map of $20\times 20$ signal and idler modes. A similar plot, but using a reduced pump power of $1\,$mW, is included in the main text. Tables \ref{table:JSI1mW} and \ref{table:JSI1p8mW} list all measured values. Furthermore, a few coincidence histograms are exhibited in SM Fig.\ref{fig:histograms}. We observe that the coincidence peak stands high over the background for energy-correlated modes; it decreases with increase in frequency detuning of either signal or idler mode, as expected.

%************ Table I. *************************************
%
\begin{table*}[!h]
\caption{Second order cross-correlation coefficients g$_{s,i}^{(2)}$ between a subset of 20x20 energy-correlated (in red) and uncorrelated spectral modes at a pump power of $1\,$ mW. We denote the $n$th signal (idler) mode by $m^{n}_{s}$ ($m^{n}_{i}$). This data is visually represented in Fig.\,3 in the main text.}

\begin{tabular}{|c||c|c|c|c|c|c|c|c|c|c|c|c|c|c|c|c|c|c|c|c|}
\hline 
 & m$_{i}^{1}$ & m$_{i}^{2}$ & m$_{i}^{3}$ & m$_{i}^{4}$ & m$_{i}^{5}$ & m$_{i}^{6}$ & m$_{i}^{7}$ & m$_{i}^{8}$ & m$_{i}^{9}$ & m$_{i}^{10}$ & m$_{i}^{11}$ & m$_{i}^{12}$ & m$_{i}^{13}$ & m$_{i}^{14}$ & m$_{i}^{15}$ & m$_{i}^{16}$ & m$_{i}^{17}$ & m$_{i}^{18}$ & m$_{i}^{19}$ & m$_{i}^{20}$\tabularnewline
\hline 
\hline
m$_{s}^{1}$ & \textit{\textcolor{red}{10.24}} & 6.2 & - & 1.43 & - & 1.12 & - & - & - & - & - & - & - & - & - & - & - & - & - & -\tabularnewline
\hline 
m$_{s}^{2}$ & 4.94 & \textit{\textcolor{red}{9.06}} & 4.25 & - & 1.46 & - & - & - & - & - & - & - & - & - & - & - & - & - & - & -\tabularnewline
\hline 
m$_{s}^{3}$ & 1.65 & 3.47 & \textit{\textcolor{red}{9.55}} & - & - & 1.36 & - & - & - & - & - & - & - & - & - & - & - & - & - & -\tabularnewline
\hline 
m$_{s}^{4}$ & - & - & 3.59 & \textit{\textcolor{red}{11.27}} & 6.77 & - & - & - & - & - & - & - & - & - & - & - & - & - & - & -\tabularnewline
\hline 
m$_{s}^{5}$ & 1.06 & - & 1.74 & - & \textit{\textcolor{red}{11.74}} & - & - & - & - & - & - & - & - & - & - & - & - & - & - & -\tabularnewline
\hline 
m$_{s}^{6}$ & - & 1.03 & - & 1.81 & - & \textit{\textcolor{red}{10.07}} & - & - & - & - & - & - & - & - & - & - & - & - & - & -\tabularnewline
\hline 
m$_{s}^{7}$ & - & - & - & - & 1.85 & - & \textit{\textcolor{red}{12.19}} & - & 3.48 & - & - & - & - & - & - & - & - & - & - & -\tabularnewline
\hline 
m$_{s}^{8}$ & - & - & - & - & - & 1.71 & - & \textit{\textcolor{red}{10.94}} & 5.84 & - & - & - & - & - & - & - & - & - & - & -\tabularnewline
\hline 
m$_{s}^{9}$ & - & - & - & - & 1.14 & - & 1.62 & - & \textit{\textcolor{red}{10.63}} & - & - & - & - & - & - & - & - & - & - & -\tabularnewline
\hline 
m$_{s}^{10}$ & - & - & - & - & - & - & - & - & - & \textit{\textcolor{red}{10.93}} & 7.52 & 1.85 & 1.45 & 1.14 & 1.23 & 1.07 & - & - & - & -\tabularnewline
\hline 
m$_{s}^{11}$ & - & - & - & - & - & - & - & - & - & 7.99 & \textit{\textcolor{red}{10.45}} & 3.71 & 1.89 & 1.27 & 1.26 & 1.01 & - & - & - & -\tabularnewline
\hline 
m$_{s}^{12}$ & - & - & - & - & - & - & - & - & - & 4.52 & 6.48 & \textit{\textcolor{red}{9.62}} & 2.95 & 2.08 & 1.43 & 1.34 & - & - & - & -\tabularnewline
\hline 
m$_{s}^{13}$ & - & - & - & - & - & - & - & - & - & 1.97 & 3.64 & 5.36 & \textit{\textcolor{red}{10.32}} & 5.11 & 2.16 & 1.46 & - & - & - & -\tabularnewline
\hline 
m$_{s}^{14}$ & - & - & - & - & - & - & - & - & - & 1.54 & 1.6 & 3.59 & 5.64 & \textit{\textcolor{red}{11.34}} & 6.47 & 2.27 & - & - & - & -\tabularnewline
\hline 
m$_{s}^{15}$ & - & - & - & - & - & - & - & - & - & 1.06 & 1.14 & 1.28 & 4.13 & 6.05 & \textit{\textcolor{red}{12.01}} & 5.81 & - & - & - & -\tabularnewline
\hline 
m$_{s}^{16}$ & - & - & - & - & - & - & - & - & - & 1.12 & 1.25 & 1.24 & 1.89 & 3.42 & 8.38 & \textit{\textcolor{red}{11.58}} & - & - & - & -\tabularnewline
\hline 
m$_{s}^{17}$ & - & - & - & - & - & - & - & - & - & - & - & - & - & 1.25 & 1.36 & - & \textit{\textcolor{red}{10.69}} & - & - & -\tabularnewline
\hline 
m$_{s}^{18}$ & - & - & - & - & - & - & - & - & - & - & - & - & - & 1.12 & - & 1.94 & - & \textit{\textcolor{red}{10.76}} & - & -\tabularnewline
\hline 
m$_{s}^{19}$ & - & - & - & - & - & - & - & - & - & - & - & - & - & - & - & - & - & - & \textit{\textcolor{red}{9.99}} & 7.43\tabularnewline
\hline 
m$_{s}^{20}$ & - & - & - & - & - & - & - & - & - & - & - & - & - & - & - & - & - & - & 3.63 & \textit{\textcolor{red}{11.64}}\tabularnewline
\hline 
\end{tabular}

\label{table:JSI1mW}

\end{table*}
%***************End Table i ******************************
%
%************** Table II. ********************************

\begin{table*}
\caption{Second order cross-correlation coefficients g$_{s,i}^{(2)}$ between a subset of 20x20 energy-correlated (in red) and uncorrelated spectral modes at a pump power of $1.8\,$mW. We denote the $n$th signal (idler) mode by $m^{n}_{s}$ ($m^{n}_{i}$)}.

	\begin{tabular}{|c||c|c|c|c|c|c|c|c|c|c|c|c|c|c|c|c|c|c|c|c|}
		\hline 
		& m$_{i}^{1}$ & m$_{i}^{2}$ & m$_{i}^{3}$ & m$_{i}^{4}$ & m$_{i}^{5}$ & m$_{i}^{6}$ & m$_{i}^{7}$ & m$_{i}^{8}$ & m$_{i}^{9}$ & m$_{i}^{10}$ & m$_{i}^{11}$ & m$_{i}^{12}$ & m$_{i}^{13}$ & m$_{i}^{14}$ & m$_{i}^{15}$ & m$_{i}^{16}$ & m$_{i}^{17}$ & m$_{i}^{18}$ & m$_{i}^{19}$ & m$_{i}^{20}$\tabularnewline
		\hline 
		\hline
		m$_{s}^{1}$ & \textit{\textcolor{red}{5.79}} & 4.33 & - & 1.4 & - & 1.01 & - & - & - & - & - & - & - & - & - & - & - & - & - & -\tabularnewline
		\hline 
		m$_{s}^{2}$ & 3.55 & \textit{\textcolor{red}{6.17}} & 3.26 & 2.05 & 1.0 & - & - & - & - & - & - & - & - & - & - & - & - & - & - & -\tabularnewline
		\hline 
		m$_{s}^{3}$ & 1.48 & 2.50 & \textit{\textcolor{red}{6.28}} & - & - & 1.17 & - & - & - & - & - & - & - & - & - & - & - & - & - & -\tabularnewline
		\hline 
		m$_{s}^{4}$ & - & - & 2.56 & \textit{\textcolor{red}{6.83}} & 4.31 & - & - & - & - & - & - & - & - & - & - & - & - & - & - & -\tabularnewline
		\hline 
		m$_{s}^{5}$ & 1.07 & - & 1.46 & - & \textit{\textcolor{red}{7.22}} & - & - & - & - & - & - & - & - & - & - & - & - & - & - & -\tabularnewline
		\hline 
		m$_{s}^{6}$ & - & 1.13 & - & 1.55 & - & \textit{\textcolor{red}{6.43}} & - & - & - & - & - & - & - & - & - & - & - & - & - & -\tabularnewline
		\hline 
		m$_{s}^{7}$ & - & - & - & - & 1.47 & - & \textit{\textcolor{red}{6.82}} & - & 3.03 & - & - & - & - & - & - & - & - & - & - & -\tabularnewline
		\hline 
		m$_{s}^{8}$ & - & - & - & - & - & 1.34 & - & \textit{\textcolor{red}{6.98}} & 4.27 & - & - & - & - & - & - & - & - & - & - & -\tabularnewline
		\hline 
		m$_{s}^{9}$ & - & - & - & - & 1.01 & - & 1.48 & - & \textit{\textcolor{red}{7.15}} & - & - & - & - & - & - & - & - & - & - & -\tabularnewline
		\hline 
		m$_{s}^{10}$ & - & - & - & - & - & - & - & - & - & \textit{\textcolor{red}{6.89}} & 5.03 & 3.56 & 1.76 & 1.23 & 1.14 & 1.07 & - & - & - & -\tabularnewline
		\hline 
		m$_{s}^{11}$ & - & - & - & - & - & - & - & - & - & 5.49 & \textit{\textcolor{red}{6.52}} & 4.44 & 2.85 & 1.19 & 1.15 & 1.11 & - & - & - & -\tabularnewline
		\hline 
		m$_{s}^{12}$ & - & - & - & - & - & - & - & - & - & 1.47 & 2.49 & \textit{\textcolor{red}{6.51}} & 3.60 & 3.17 & 1.14 & 1.23 & - & - & - & -\tabularnewline
		\hline 
		m$_{s}^{13}$ & - & - & - & - & - & - & - & - & - & 1.14 & 1.48 & 2.46 & \textit{\textcolor{red}{6.89}} & 4.01 & 2.96 & 1.3 & - & - & - & -\tabularnewline
		\hline 
		m$_{s}^{14}$ & - & - & - & - & - & - & - & - & - & 1.18 & 1.18 & 1.57 & 3.07 & \textit{\textcolor{red}{7.16}} & 3.94 & 2.59 & - & - & - & -\tabularnewline
		\hline 
		m$_{s}^{15}$ & - & - & - & - & - & - & - & - & - & 1.11 & 1.21 & 1.35 & 1.74 & 4.12 & \textit{\textcolor{red}{7.38}} & 5.88 & - & - & - & -\tabularnewline
		\hline 
		m$_{s}^{16}$ & - & - & - & - & - & - & - & - & - & 1.19 & 1.14 & 1.23 & 1.42 & 1.79 & 4.54 & \textit{\textcolor{red}{7.64}} & - & - & - & -\tabularnewline
		\hline 
		m$_{s}^{17}$ & - & - & - & - & - & - & - & - & - & - & - & - & - & 1.17 & 1.39 & - & \textit{\textcolor{red}{7.44}} & - & - & -\tabularnewline
		\hline 
		m$_{s}^{18}$ & - & - & - & - & - & - & - & - & - & - & - & - & - & 1.29 & - & 1.86 & - & \textit{\textcolor{red}{7.60}} & - & -\tabularnewline
		\hline 
		m$_{s}^{19}$ & - & - & - & - & - & - & - & - & - & - & - & - & - & - & - & - & - & - & \textit{\textcolor{red}{8.00}} & 5.26\tabularnewline
		\hline 
		m$_{s}^{20}$ & - & - & - & - & - & - & - & - & - & - & - & - & - & - & - & - & - & - & 2.84 & \textit{\textcolor{red}{6.41}}\tabularnewline
		\hline 
	\end{tabular}
	
\label{table:JSI1p8mW}
\end{table*}

%%%%%%%%%%%%%%%%%%%%%%%%%%%%%%%%%%%%%%%%%%%%%%%%%%%%%%%%%%%%%%%%%%%%%%%%

\section{Simultaneous collection of the idler modes: demultiplexer design and implementation}

According to the frequency multiplexed quantum repeater scheme discussed in SM Section I (see SM Fig.\ref{fig:Repeater}), the Bell state measurements between photons must be performed in frequency-resolved fashion. This requires simultaneous collection of individual spectrally demultiplexed photon modes in distinct spatial channels. To achieve this, we design and implement an optical VIPA setup integrated with an array of optical fibers. 

 The fiber array and the collection lens system are designed based on two main requirements: (1) the focal spot size produced by the selected focusing lens (or a system of lenses) in the focal plane needs to match both the mode size and the NA of the fiber mode to ensure high coupling efficiency, and (2) the spacing between focal spots corresponding to  neighbouring idler modes needs to be equal to the distance between the core centers of the adjacent fibers in the array. This last condition imposes a lower bound on the acceptable spatial separation between neighbouring spectral channels in the focal plane, and is due to the minimum cladding diameter for a single mode fiber.%, \textcolor{red}{(I do not understand the following, isn't the cladding diameter enough? Can the following be rephrased or removed (in case it is not needed)?) both from a technically achievable perspective and as dictated by the requirements of good light confinement and low cross-talk between adjacent fiber channels).} \ALT{\bf The cladding diameter of a standard fiber is 125 micros. The claddings of some fibers were etched down to ~82 microns to make this array possible. They could etch them further down, but 1- at some point light is not very well confined and x-talk increases, and 2- handling becomes very tricky. But yes, we can remove this text. Fixed - commented it out.}
 
 %   OLD:
 %First, we determine the optimized focal length of the lens behind the VIPA. Mainly, we check the beam waist sizes (at 1532 nm) for different lenses. Lenses with long focal length (e.g. 1000 mm) focus the modes into a large spot size and create large spatial separation ($\sim$mm) between the modes, which makes it easy to measure the intensity distribution. However, in this case the NA being too low, an efficient coupling to the fiber cannot be achieved. For instance, a combination of input beam width of 8 mm and a focusing lens with focal length of 1000 mm yields an NA of 0.004 and 0.001 along the mapping and vertical directions, respectively. Lenses with shorter focal lengths yield higher NA, which helps in coupling the modes to the optical fibers efficiently. 

\begin{figure}[b]
\includegraphics[width=0.55\linewidth]{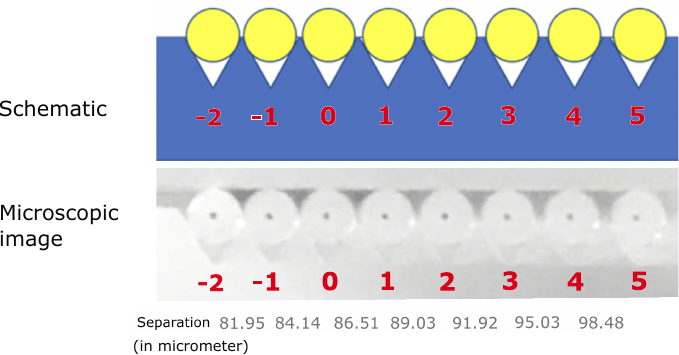}
\caption{Schematic of the spatial distributions of the spectral modes spaced by $6.5\,$GHz together with a microscopy image of the designed fiber array.}
\label{fig:Modes8_fiber_array}
\end{figure}

 As described above, a lens with focal length of $45\,$mm provides optimal matching of a VIPA output mode to the mode of an SMF-28 fiber (in terms of both the spot size and the NA). However, due to tight focusing, the spatial distances between two adjacent spatial modes separated by $6.5\,$GHz, are in the range of $61-72\,\mu$m (as shown in Fig. \ref{fig:VIPA_charctrzn}(a), the spatial mode spacing varies along the VIPA profile). 
 Such small distances make it difficult to accommodate the cladding layers between two adjacent fiber cores while preserving low cross-talk between them. %This would compromise simultaneous coupling of the multiple spectral modes of the photons to an array of fibers and would introduces considerable cross-talks between adjacent modes. 
 Instead, we find that a focusing lens, or a system of lenses, with (effective) focal length of $60\,$mm is optimal for our purpose. Indeed, it results in spatial separations in the range of $81 \,-\,99\,\mu$m between the adjacent modes and provides an NA$=\,0.21$(calculated for an f$=$60 mm lens of 1 inch diameter), which allows us to optimally couple the spectral modes into SMF-28 fibers. We simulate the mapping positions of 8 spectral modes at $1532\,$nm separated by $6.5\,$GHz (equal to the idler mode spacing) for a focusing lens with focal length of $60\,$mm, and design and prepare (manufactured by PHIX Photonics Assembly, Netherlands) an array of 8 fibers, which are spaced by distances similar to the ones obtained from the simulation. SM Fig. \ref{fig:Modes8_fiber_array} shows a microscopy image of the prepared fiber array and a comparison between the measured and simulated inter-core spacings. In order to control the spacing between the VIPA modes and optimally match their mapping positions to the corresponding fiber cores, we use a telescopic system comprising two achromatic doublets with identical focal lengths of $100\,$mm. This telescope has an effective focal length of $60\,$mm (the separation between the two doublets is $28.37\,$mm and the working distance is $33.64\,$mm), and can correct for small deviations in the mode spacing by offering a freedom to fine-tune the focal distance.

\begin{figure}[b]
\includegraphics[width=0.65\linewidth]{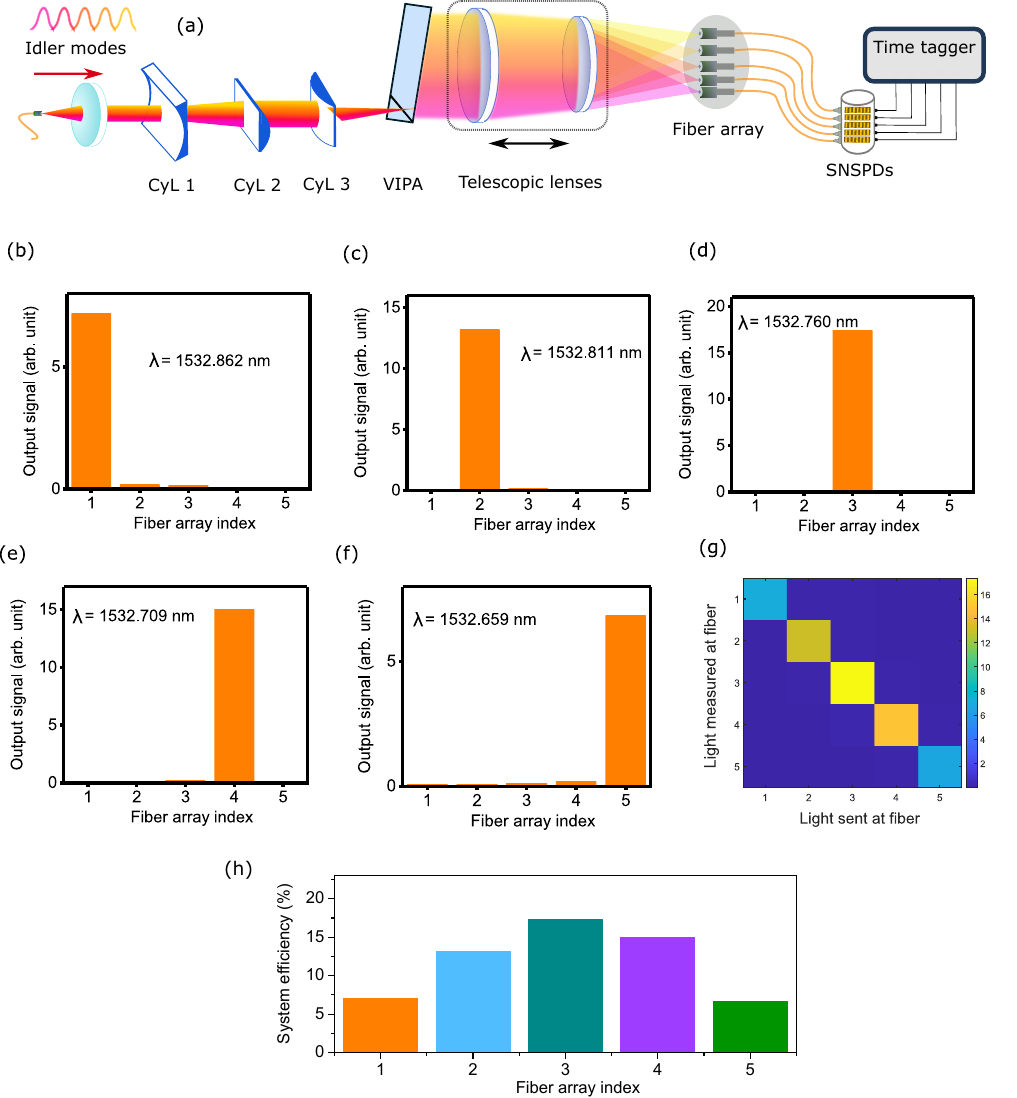}
\caption{(a) Optical setup for optimally coupling spectral modes separated by ${6.5}$ GHz to the designed fiber array. (b) - (f): cross-talk for five spectral modes between different fiber channels. (g) Summary of panels (b)-(f). (h) System efficiencies for the spectral modes collected through the five fiber channels used in our experiments.
\label{fig:Array_couplings}}
\end{figure}

%\begin{figure}
%\includegraphics[width=0.50\linewidth]{syst_effncies.pdf}
%\caption{System efficiencies for the spectral modes collected through the five fiber channels used in our experiments.  \ALT{{\bf Tanmoy}, could you please include this figure into the previous one as an (h) panel? I know it's crowded, but it's all about the setup and its performance characterization.}\TCcomment{Done}.
%\label{fig:syst_effncies}}
%\end{figure}

The experimental VIPA-based setup for demultiplexing and simultaneously collecting spectral modes using our fiber array is shown in Fig. \ref{fig:Array_couplings}. To achieve efficient coupling, we mount the fiber array on a tip-tilt with rotation stage, which is mounted on a NanoMax flexure stage (Thorlabs, Inc.) that provides 3 additional translational degrees of freedom (X, Y, Z). For our proof-of-principle demonstration, we couple light to 5 fibres and proceed with the frequency-resolved measurements. The fibres are indexed with $n\,=\,1$ to $5$. We use a laser to generate spectral modes at ${\lambda}\,=\,1532.608$nm$\,+\,n\,\times\,6.5\,$GHz with $n\,=\,1$ to $5$, where ${\lambda}_3$ spectrally aligns along the peak of the intensity profile shown in SM Fig. \ref{VIPA-9freqs}. Light at wavelength ${\lambda}_n$ is collected through the corresponding  fiber of the array. After optimizing the fiber array positions simultaneously for all the five wavelengths using the 6-axis control stage (three translational and three rotational degrees of freedom), we determine inter-channel cross-talks by sending light at wavelengths ${\lambda}_1$, ${\lambda}_2$, ${\lambda}_3$, ${\lambda}_4$ and ${\lambda}_5$ and by measuring optical power at the output of each of the 5 channels. SM Figs. \ref{fig:Array_couplings}(b) to (g) show that the light at each specific wavelength is transmitted almost elusively though the corresponding spatial (fiber) channel, with almost negligible signal passing through the neighboring channels. We observe an inter-channel cross-talk $<\,45\,$dB for our fiber array system. %This ensures that our fiber array does not introduce any spectral impurity while collecting photons of each of the individual spectral idler modes through the fiber channels. 
We measure a system efficiency (from the input fiber through the VIPA and demultiplexing setup into the collecting fiber) of 17$\%$ for the $3^{rd}$ fiber, which reduces to approximately 8$\%$ on both sides for fibers number 1 and 5, as shown in Fig. \ref{fig:Array_couplings}(h).%{fig:syst_effncies}.

\begin{figure}[h!]
\includegraphics[width=0.8\linewidth]{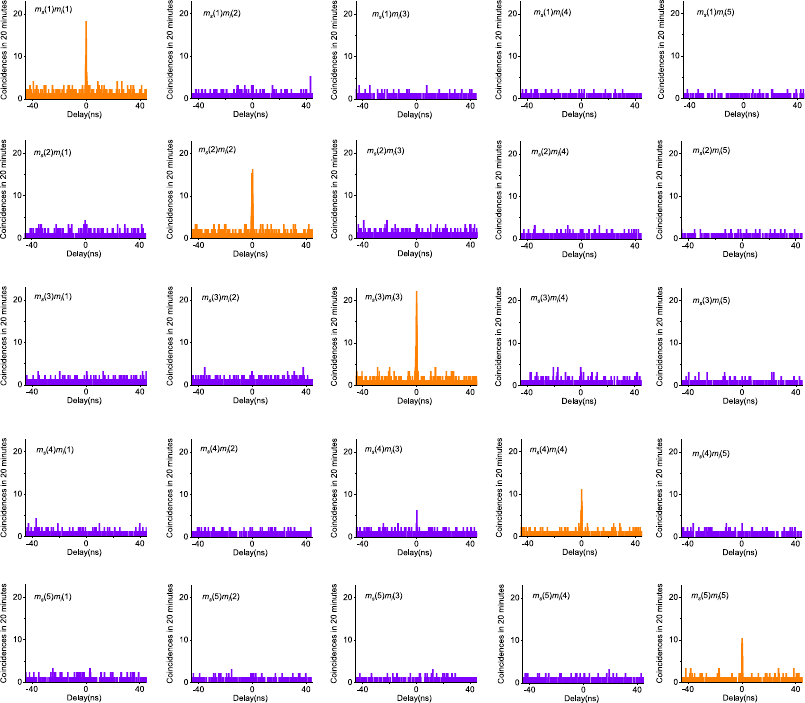}
\caption{Coincidence histograms between $5\times 5$ spectral modes of the signal and idler photons. Idler modes are demultiplxed using the VIPA and collected in different spatial channels using the fiber array. Signal modes are spectrally filtered using a spectral trench prepared in Tm$^{3+}$:LiNbO$_3$.
\label{fig:All_coincidences}}
\end{figure}
 
%%%%%%%%%%%%%%%%%%%%%%%  Section X. %%%%%%%%%%%%%%%%%%%%%%%%%%%%%%%%%%

\section{Spectral filtering of signal modes: burning wide spectral holes in T\MakeLowercase{m}$^{3+}$:L\MakeLowercase{i}N\MakeLowercase{b}O$_3$}

As a first step towards creating a multimode atomic frequency comb (AFC)-based quantum memory that allows storing the spectral modes of the signal photons, we burn spectral holes (trenches) of 100 MHz width into the inhomogeneously broadened 795 nm $^{3}H_{6}$ to $^{3}H_{4}$ absorption line of Tm$^{3+}$:LiNbO$_3$ (see Fig. 3(c) in the main text). The trenches match the spectral positions of the signal photon modes. We apply a magnetic field of 400 Gauss along the c-axis of the crystal and employ one of the resulting nuclear spin sub-level of $^{3}H_{4}$ (with lifetime $>1\,$s) as a shelving state for persistent hole burning. The experiments are performed at a temperature of $600\,$mK, and the power of the laser used for hole burning is $300\,\mu$W before the Tm$^{3+}$:LiNbO$_3$ crystal. As shown in  Fig. 3(b) of the main text, the burn duration is $50\,$ms and the measurement time per cycle is $150\,$ms. Since the excited state lifetime is $160\,\mu$s, we wait $2\,$ms between hole burning and measurement to ensure that no population remains in the $^3$H$_4$ level and no spontaneously emitted photons can mask the photons created by the SPDC source.

%%%%%%%%%%%%%%%%%%%%%%  Section XI  %%%%%%%%%%%%%%%%%%%%%%%%%

\section{Frequency-resolved cross-correlation measurements}

To perform joint spectral measurements between demultiplexed idler modes and spectrally filtered signal modes, we use 5 fiber array channels to collect 5 idler modes, demultiplexed by the VIPA. We first identify the energy-correlated signal modes using the FP$_{signal}$ cavity filter as described in section SM VIII. However, since the linewidth of the FP$_{signal}$ is $6.1\,$GHz, it also weakly transmits the two adjacent signal modes, which results in excessive g$^{(2)}_{s,i}$ values for non-energy-correlated modes. To add additional spectral selectivity, the FP$_{signal}$-filtered signal modes additionally pass through the 100 MHz wide spectral trench burned into the transmission line of Tm$^{3+}:$LiNbO$_3$, which significantly reduces the undesired (adjacent) modes. Coincidence histograms are recorded using a time-tagger (Swabian Instruments GmbH). The frequency-resolved signal-idler optical interface is schematically shown in Fig.1(a) of the main text. 

\begin{figure}[h!]
\includegraphics[width=0.55\linewidth]{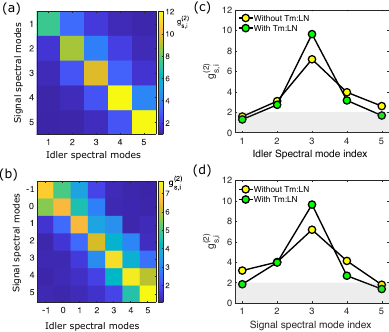}
\caption{Map of cross-correlation coefficients between spectral modes of signal and idler photons (a) for $5\times 5$ modes measured using the spectral trench within the Tm$^{3+}$:LiNbO$_3$ absorption line, and (b) for $7\times 7$ modes without the spectral trench. Variation of cross-correlation coefficients when (c) the idler mode (for signal mode 3) and (d) the signal mode (for idler mode 3) is shifted. Both panels include data with and without the Tm$^{3+}$:LiNbO$_3$ (Tm:LN) spectral filter. The grey regions denote the classical domain (g$^{(2)}_{s,i}<$ 2).
\label{fig:JSIs}}
\end{figure}

Using the coincidence histograms measured between the spectral trench-filtered signal modes and VIPA-fiber-array demultiplexed idler modes, we characterize the joint spectral properties of the photons. SM Fig. \ref{fig:All_coincidences}(a) shows 25 coincidence rates between $5\times 5$ signal-idler modes. The numerical values of the coincidence rates and of g$^{(2)}_{s,i}$ are given in table \ref{table:JSI1p8mWTmLN_fiber_array} and \ref{table:Coinc_1p8mWTmLN_fiber_array}, respectively. Fig. \ref{fig:All_coincidences}(b) shows coincidences between $7\times 7$ signal-idler modes measured using a single fiber after the VIPA for collecting the modes of 1532 nm photons and only the $6.1\,$GHz cavity filter for the $795\,$nm photons. We compare the variation of the g$^{(2)}_{s,i}$ values measured with and without the Tm$^{3}$:LiNbO$_3$ filer, when one shifts from the energy-correlated signal and idler modes to the non-correlated ones in steps, which are shown in Fig. \ref{fig:All_coincidences} (c) and (d), respectively. These plots show that inclusion of Tm$^{3}$:LiNbO$_3$ filter results in an enhanced reduction of g$^{(2)}_{s,i}$ for the no-correlated modes, which signifies that Tm$^{3}$:LiNbO$_3$ spectral filer adds additional spectral selectivity for the signal photons.

\begin{table*}[h]
\caption{Coincidence rates between a subset of $5 \times 5$ energy-correlated (in red) and uncorrelated spectral modes at a pump power of $1.8\,$mW. We denote the $n^{th}$ signal (idler) mode by $m_{s}(n)$ ($m_{i}(n)$)}.

\begin{tabular}{|c||c|c|c|c|c|}
\hline
& m$_{i}(1)$ & m$_{i}(2)$ & m$_{i}(3)$ & m$_{i}(4)$ & m$_{i}(5)$ \tabularnewline
\hline 
\hline 
m$_{s}(1)$ & \textit{\textcolor{red}{8.7083}} & 1.7917 & 0.7917 & 0.7500 & 0.2500 \tabularnewline
\hline
m$_{s}(2)$ & 2.2083 & \textit{\textcolor{red}{11.0833}} & 1.5000 & 0.5833 & 0.6667 \tabularnewline
\hline
m$_{s}(3)$ & 0.2500 & 1.4583 & \textit{\textcolor{red}{12}} & 1.8750 & 0.3333 \tabularnewline
\hline
m$_{s}(4)$ & 0.4167 & 0.5833 & 1.9583 & \textit{\textcolor{red}{5.6250}} & 0.5000 \tabularnewline
\hline
m$_{s}(5)$ & 0.2083 & 0.2500 & 0.4167 & 1 & \textit{\textcolor{red}{4.5833}} \tabularnewline
\hline
\end{tabular}

\label{table:JSI1p8mWTmLN_fiber_array}
\end{table*}

\begin{table*}[!h]
\caption{Second order cross-correlation coefficients g$_{s,i}^{(2)}$ between a subset of $5\times 5$ energy-correlated (in red) and uncorrelated spectral modes at a pump power of 1.8 mW. We denote the $n^{th}$ signal (idler) mode by $m_{s}(n)$ ($m_{i}(n)$)}.

\begin{tabular}{|c||c|c|c|c|c|}
\hline
& m$_{i}(1)$ & m$_{i}(2)$ & m$_{i}(3)$ & m$_{i}(4)$ & m$_{i}(5)$ \tabularnewline
\hline 
\hline 
m$_{s}(1)$ & \textit{\textcolor{red}{9.8030}} & 2.6741 & 1.3324 & 1.5571 & 0.5017 \tabularnewline
\hline
m$_{s}(2)$ & 4.0212 & \textit{\textcolor{red}{12.0253}} & 2.2222 & 1.2821 & 1.9704 \tabularnewline
\hline
m$_{s}(3)$ & 0.7463 & 2.1472 & \textit{\textcolor{red}{13.5466}} & 3.1960 & 0.9709 \tabularnewline
\hline
m$_{s}(4)$ & 1.5015 & 1.5152 & 3.5392 & \textit{\textcolor{red}{10.3846}} & 1.1194 \tabularnewline
\hline
m$_{s}(5)$ & 0.6684 & 0.7426 & 1.1062 & 2.0305 & \textit{\textcolor{red}{9.6491}} \tabularnewline
\hline
\end{tabular}

\label{table:Coinc_1p8mWTmLN_fiber_array}
\end{table*}

\end{document}